\documentclass[letterpaper,british,english]{emulateapj}
\usepackage{ae,aecompl}
\usepackage[T1]{fontenc}
\usepackage{xcolor}
\usepackage{verbatim}
\usepackage{textcomp}
\usepackage{amsbsy}
\usepackage{amstext}
\usepackage{amssymb}
\usepackage{graphicx}
\PassOptionsToPackage{normalem}{ulem}
\usepackage{ulem}

\makeatletter

\slugcomment{}
\shorttitle{}
\shortauthors{}

\usepackage{amsmath}

\makeatother

\usepackage{babel}
\begin{document}

\title{ATMOSPHERIC RETRIEVAL FOR SUPER-EARTHS: UNIQUELY CONSTRAINING THE
ATMOSPHERIC COMPOSITION WITH TRANSMISSION SPECTROSCOPY}

\author{Bjoern Benneke and Sara Seager}

\affil{Department of Earth, Atmospheric and Planetary Sciences, Massachusetts
Institute of Technology, Cambridge, MA 02139, USA\textit{}\\
}

\email{bbenneke@mit.edu}
\begin{abstract}
We present a retrieval method based on Bayesian analysis to infer
the atmospheric compositions and surface or cloud-top pressures from
transmission spectra of exoplanets with general compositions. In this
study, we identify what can unambiguously be determined about the
atmospheres of exoplanets from their transmission spectra by applying
the retrieval method to synthetic observations of the super-Earth
GJ~1214b. Our approach to infer constraints on atmospheric parameters
is to compute their joint and marginal posterior probability distributions
using the Markov Chain Monte Carlo technique in a parallel tempering
scheme. A new atmospheric parameterization is introduced that is applicable
to general atmospheres in which the main constituent is not known
a priori and clouds may be present.

Our main finding is that a \textit{unique} constraint of the mixing
ratios of the absorbers and two spectrally-inactive gases (such as
$\mathrm{N_{2}}$ and primordial $\mathrm{H}_{2}\mathrm{+He}$) is
possible if the observations are sufficient to quantify both (1) the
broadband transit depths in at least one absorption feature for each
absorber and (2) the slope and strength of the molecular\textit{ }Rayleigh
scattering signature. A second finding is that the surface pressure
or cloud-top pressure can be quantified if a surface or cloud deck
is present at low optical depth. A third finding is that the mean
molecular mass can be constrained by measuring either the Rayleigh
scattering slope or the shapes of the absorption features, thus enabling
one to distinguish between cloudy hydrogen-rich atmospheres and high
mean molecular mass atmospheres. We conclude, however, that without
the signature of molecular Rayleigh scattering---even with robustly
detected infrared absorption features (>10$\sigma$)---there is no
reliable way to tell from the transmission spectrum whether the absorber
is a main constituent of the atmosphere or just a minor species with
a mixing ratio of $X_{\mathrm{abs}}$<0.1\%. The retrieval method
leads us to a conceptual picture of which details in transmission
spectra are essential for unique characterizations of well-mixed exoplanet
atmospheres.
\end{abstract}


\section{Introduction\label{sec:Introduction}}

Major advances in the detection and characterization of exoplanet
atmospheres have been made over the last decade. To date, several
dozen hot Jupiter atmospheres have been observed by the \textit{Spitzer
Space Telescope}, \textit{Hubble Space Telescope} and/or ground-based
observations. Observational highlights include the detection of molecules
and atoms \citep[e.g.,][]{charbonneau_detection_2002,deming_infrared_2005,seager_exoplanet_2010}
and the identification of thermal inversion \citep{knutson_3.68.0_2008}.
Recent observational efforts \citep[e.g.,][]{bean_ground-based_2010,croll_broadband_2011,berta_flat_2012}
suggest that the continuous improvements in observational techniques
will enable us to extend the field of atmospheric characterization
to the regime of super-Earths (exoplanets with mass between 1 and
10 $M_{\bigoplus}$) in the near future.

Since super-Earth exoplanets lie in the intermediate mass range between
terrestrial planets and the gas/ice giants in the Solar System, compelling
questions arise as to the nature and formation histories of these
objects and whether they are capable of harboring life. A potential
way of answering these questions is to constrain the molecular compositions
and thicknesses of their atmospheres from spectral observations of
the transmission and/or emission spectra \citep{miller-ricci_atmospheric_2009}.
While a thick hydrogen/helium envelope would indicate that their formation
histories are similar to those of the gas or ice giant planets in
the Solar System, super-Earths that are predominately solid planets
may be scaled-up analogs of the terrestrial planets in our solar system.
Alternative scenarios of planets different in nature to the solar
system planets, such as planets mainly composed of water or carbon
compounds, have been proposed as well \citep{kuchner_volatile-rich_2003,leger_new_2004,kuchner_extrasolar_2005}.

As the first observations of the transmission spectrum of the super-Earth
GJ~1214b become available, the current practice in interpreting these
spectra is to check the observations for their agreement to preconceived
atmospheric scenarios \citep{miller-ricci_nature_2010,bean_ground-based_2010,croll_broadband_2011}.
There are two dangers with this approach: First, even if a good fit
is reached between the data and the model spectrum of a preconceived
scenario, we do not know whether we actually understand the nature
of the planet or whether we have simply found one out of several possible
scenarios matching the data. Second, and even more important, we will
not be able to understand planets that do not fit our preconceived
ideas. These planets, however, would likely represent the most compelling
science cases as they may provide new insights into planetary formation
and evolution, atmospheric chemistry, or astrobiology.

Here, we present a new tool for the interpretation of transmission
spectra of transiting super-Earth and mini-Neptune exoplanets. The
approach is fundamentally different from previous work on super-Earth
atmospheres in that we retrieve constraints on the atmospheric composition
by exploring a wide range of atmospheres with self-consistent temperature
structures. Our approach builds on the idea introduced in the pioneering
works on hot Jupiters by \citet{madhusudhan_temperature_2009} and
\citet{madhusudhan_carbon-rich_2011} to use Monte Carlo methods to
explore the parameter space for solutions that are in agreement with
the observations. The method presented here is different in three
ways. First, our retrieval method is applicable to atmospheres of
general composition and considers the presence of a cloud deck or
solid surface. We introduce concepts of compositional data analysis,
a subfield of statistical analysis, to treat the mixing ratios of
all molecular constituents equally while ensuring that the sum of
the mixing ratios is unity. Second, we use a radiative-convective
model to calculate a temperature profile that is self-consistent with
the molecular composition of each model atmosphere. Third, we conduct
a full Bayesian analysis and infer our constraints on atmospheric
parameters directly by marginalizing the joint posterior probability
distribution obtained from the Markov Chain Monte Carlo (MCMC) simulation.
We therefore obtain the most-likely estimates and statistically significant
Bayesian credible intervals for each parameter. \citet{madhusudhan_high_2011}
used the MCMC algorithm to explore the model parameter space in the
search for regions that provide good fits to the data. Based on the
parameter exploration, they were able to report contours of constant
goodness-of-fit in the parameters space. Contours of constant goodness-of-fit,
however, cannot directly be related to the confidence regions of the
desired parameters. 

The retrieval method presented in this work is different from optimum
estimation retrieval, as described by \citet{rodgers_inverse_2000}
and recently applied to exoplanets by \citet{lee_optimal_2011} and
\citet{line_information_2012}, in that we derive the full probability
distributions and Bayesian credible regions for the desired atmospheric
parameters, while optimum estimation retrieval assumes Gaussian errors
around a single best-fitting solution. Highly non-Gaussian uncertainties
of the atmospheric parameters need to be considered for noisy exoplanet
observations because the observable atmospheric spectra are highly
nonlinear functions of the desired atmospheric parameters, and a large
volume in the parameter space are generally compatible with noisy
exoplanet spectra. Our approach to calculate the joint posterior probability
distribution using MCMC is computationally intensive ($\sim10^{5}$
model evaluations are required), but it enables us to extract all
that can be inferred about the atmospheric parameters from the observational
data. The uncertainty of individual atmospheric parameters introduced
by complex, non-Gaussian correlations to other parameters is accounted
for in a straightforward way by marginalizing over the remaining parameters.
Optimum estimation retrieval, in contrast, searches for a single best-fitting
solution using the Levenberg-Marquardt algorithm. Gaussian uncertainties
are estimated around the best-fitting solution by linear analysis
\citep{rodgers_inverse_2000,lee_optimal_2011} or by performing multiple
retrieval runs with individual parameters fixed at particular values
\citep{lee_optimal_2011}. Optimum estimation retrieval requires fewer
model evaluations (typically $\sim10-20$ per retrieval run, multiplied
by the number of retrieval runs performed with individual parameters
fixed) and may therefore allow the use of more complex atmospheric
models given the same computational resources. For noisy exoplanet
spectra, however, the optimal estimation retrieval may not correctly
represent the confidence regions of the atmospheric parameters because
the uncertainties of the atmospheric parameters are highly non-Gaussian.

In this work, we investigate what we can learn about the atmospheres
of super-Earths solely based on transmission spectroscopy by applying
the retrieval method to a sample of synthetic observations of different
super-Earth scenarios. Previous studies have shown that the atmospheric
composition as well as the presence of a solid surface and clouds
affect the planetary spectrum \citep[e.g.,][]{seager_theoretical_2000,des_marais_remote_2002,ehrenreich_transmission_2006},
but no comprehensive study of the degeneracy of these effects has
been performed. As a result, it is not fully understood which individual
atmospheric parameters can be inferred \textit{uniquely} from the
spectrum and which parameters are strongly correlated or degenerate.
In particular, for super-Earth atmospheres for which the formation
history and subsequent evolution is not understood, we do not know
the mean molecular mass and the thickness of the atmosphere a priori.
For such planets, the depths of individual absorption features in
the spectrum are affected not only by the mixing ratio of the absorbing
molecular species, but also by the unknown mean molecular mass of
the background atmosphere and the surface or cloud deck pressure.
Strong correlations or degeneracies between these atmospheric properties
are therefore expected, but have not been addressed sufficiently in
the literature.

The paper outline is as follows. We introduce the new retrieval method
in Section 2. Section 3 describes the synthetic observations of the
super-Earth scenarios. In Section 4, we introduce a conceptual picture
of the information contained in transmission spectra and present numerical
results. Section 5 discusses the overall approach to obtain atmospheric
constraints from observations and expands on the effect of hazes and
stratified atmospheres. We also discuss a new way of planning observations
using our atmospheric retrieval method, and comment on the complementarity
between atmospheric retrieval and self-consistent modeling of atmospheres.
In Section 6, we present a summary of our results and the conclusions.

\section{Method}

Our eventual aim is to characterize the atmospheres of exoplanets
based on observations of their transmission spectra and without prior
knowledge of their natures. The primary inputs to the retrieval method
are observations of the wavelength-dependent transit depth during
the primary transit. The outputs are the best estimates and confidence
regions of the desired atmospheric properties, such as the mixing
ratios of the molecular constituents and the surface/cloud-top pressure.
We solve the \textquotedbl{}inverse\textquotedbl{} problem to regular
atmospheric modeling, in which the transmission spectrum is calculated
given a description of the composition and state of the atmosphere. 

The essential part of defining the retrieval problem is to specify
a set of parameters that both unambiguously defines the state of atmospheres
and may be constrained by transit observations. We employ an atmospheric
\textquotedbl{}forward\textquotedbl{} model to represent the physical
relation between the set of atmospheric parameters and the observable
transit depths. Given a set of observations, we retrieve constraints
on atmospheric parameters by performing a Bayesian analysis using
the atmospheric forward model and the MCMC technique. The joint posterior
probability distribution provided by the MCMC simulation represents
the complete state of knowledge about the atmospheric parameters in
the light of the observational data.

\subsection{Atmosphere Parameterization\label{sub:Parametrization}}

We propose a parametric description of the atmosphere guided by the
information available in exoplanet transmission spectra. Our approach
is to treat the atmosphere near the terminator as a well-mixed, one-dimensional
atmosphere and describe the unknown molecular composition, thickness,
and albedo of this atmosphere by free parameters. The motivation for
treating the atmosphere as well-mixed is to keep the number of parameters
to a minimum to avoid overfitting of the sparse data available in
the near future, while ensuring that all atmospheric properties that
considerably affect the retrieval from the spectrum are described
by free parameters. For atmospheres with a stratified composition,
the retrieval method determines an altitude-averaged mixing ratio
that best matches the observed transmission spectrum (Section \ref{sub:Well-mixed-Atmosphere}).

The unknown temperature profile at the terminator presents a challenge.
While the pressure dependence of the temperature profile has only
a secondary effect on the transmission spectrum and can likely not
be retrieved given that the molecular composition is unknown a priori,
the temperature does affect the scale height and may affect the constraints
on other parameters. Our approach is not to retrieve the temperature
profile, but to account for the uncertainty introduced by the unknown
temperature on the retrieved composition and surface pressure. We
therefore introduce a free parameter for the planetary albedo and
calculate the temperature profile consistent with the molecular composition
and the planetary albedo for each model atmosphere. In the MCMC analysis,
the albedo is then allowed to vary over the range of plausible planetary
albedos. Marginalizing the posterior distribution over all albedo
values allows us to account for the uncertainty of the composition
and surface pressure introduced by the unknown albedo.

Our proposed model has the following free parameters.

\paragraph{\textit{Volume mixing ratios of atmospheric constituents }}

We parameterize the composition of the atmosphere by the volume mixing
ratios of all plausibly present molecular species. The volume mixing
ratio $X_{i}$ (or equivalently the mole fraction) is defined as the
number density of the constituent $n_{i}$ divided by the total number
density of all constituents in the gas mixture $n_{tot}$. No assumptions
on the elemental composition, chemistry, or formation and evolution
arguments are made. In contrast to the work on hot Jupiters by \citet{madhusudhan_temperature_2009},
we cannot assume a hydrogen-dominated atmosphere. We therefore reparameterize
the mixing ratios with the centered-log-ratio transformations described
in Section \ref{par:Compositional-Analysis}. The transformation ensures
that all molecular species are treated equally and no modification
is required when applying the retrieval method to atmospheres with
different main constituents.

\paragraph{\textit{Surface or cloud deck pressure}}

We introduce the ``surface'' pressure $P_{\mathrm{surf}}$ as a free
parameter, where the surface is either the ground or an opaque cloud
deck. Solid surfaces and opaque cloud decks have the same effect on
the transmission spectrum and we cannot discriminate between them.
Our parameterization of the surface is applicable for rocky planets
with a thin atmosphere as well as planets with a thick gas envelope.
For thick atmosphere, for which there is no surface affecting the
transmission spectrum, the inference of the surface pressure parameter
provides a lower bound on the thickness of the cloud-free part of
the atmosphere.

\paragraph{\textit{Planet-to-star radius ratio parameter}}

We define the planet-to-star radius ratio parameter, $R_{P,10}/R_{*}$,
as the planetary radius at the 10 mbar pressure level, $R_{P,10}$,
divided by the radius of the star $R_{*}$. Our approach to define
the planetary radius at a fixed pressure level rather than at the
surface avoids degeneracy between the planetary radius and the surface
pressure for optically thick atmospheres for which the surface pressure
cannot be constrained. It enables us to perform the retrieval for
all types of planets without knowing a priori whether or not the planet
has a surface. For planets with a surface pressure lower than 10 mbar,
we still model an atmosphere down to the 10 mbar level and consider
layers at pressure levels with $P>P_{\mathrm{surf}}$ to be opaque.

\paragraph{\textit{Planetary albedo}}

While the goal is not to infer the planetary albedo, $A_{p}$, we
wish to account for the uncertainty in the retrieved mixing ratios
and surface pressure introduced by the unknown planetary albedo and
equilibrium temperature. We therefore define the albedo as a free-floating
parameter and assign a prior to the albedo parameter that reflects
our ignorance of the albedo.

\paragraph{\textit{Fixed input parameters}}

Additional input parameters that are fixed in this study are the radius
of the star, $R_{*}$, the planetary mass known from radial velocity
measurements, $M_{p}$, and the semi-major axis of the planet's orbit,
$a_{p}$. The effect of the uncertainties associated with these parameters
on the retrieval results may be accounted for by letting the parameters
float and assigning them a prior distribution.

\subsection{Atmospheric \textquotedbl{}Forward\textquotedbl{} Model \label{sub:Atmosphere-"Forward"-Model}}

The objective of the atmospheric \textquotedbl{}forward\textquotedbl{}
model is to generate transmission spectra for a wide range of different
atmospheric compositions and thicknesses. Given a set of atmospheric
parameters (Section \ref{sub:Parametrization}), our model uses line-by-line
radiative transfer in local thermodynamic equilibrium, hydrostatic
equilibrium, and a temperature profile consistent with the molecular
composition to determine the transmission spectrum. The output of
each model run is a high-spectral resolution transmission spectrum
as well as simulated instrument outputs given the response functions
of the instrument channels used in the observations. To obtain convergence
of the posterior probability distribution in the MCMC inference, the
model must efficiently generate $\sim10^{5}$ atmospheric model spectra.

\subsubsection{Opacities \label{sub:Opacities}}

\paragraph{\textit{Molecular absorption}}

We determine the monochromatic molecular absorption cross sections
from the HITRAN database \citep{rothman_hitran_2009} below 800 K.
At temperatures higher than 800K we account for the high-temperature
transitions of the gases $\mathrm{H_{2}O}$, $\mathrm{CO_{2}}$, and
$\mathrm{CO}$ using the HITEMP database \citep{rothman_hitemp_2010}.
We account for $\mathrm{H}_{2}$-$\mathrm{H_{2}}$ collision-induced
absorption using opacities from \citet{borysow_collision-induced_2002}.

To speed up the evaluation of a large number of atmospheric models,
we first determine the wavelength-dependent molecular cross sections
for each of the considered molecular species on a temperature and
log-pressure grid and then interpolate the cross section for the required
conditions. In the upper atmosphere, spectral lines become increasingly
narrow, requiring a very high spectral resolution to exactly capture
the shapes of the thin Doppler-broadened lines \citep{goody_atmospheric_1995}.
Instead of ensuring that each line shape at low pressure is represented
exactly, we choose an appropriate spectral resolution for the line-by-line
simulation by ensuring that the simulated observations are not altered
by more than 1\% of the observational error-bar when the spectral
resolution is doubled or quadrupled. While there are many methods
proposed in the literature to reduce the computation time (e.g., correlated-\textit{k}
methods and band-models; \citealp{goody_atmospheric_1995}), the accuracy
of such methods is hard to assess when the atmospheric composition
is completely unknown a priori.

\paragraph{\textit{Rayleigh scattering} }

The Rayleigh scattering cross section, $\sigma_{R,i}$, for a molecular
species $i$ can be expressed in cgs units as

\begin{equation}
\sigma_{R,i}\left(\nu\right)=\frac{24\pi\nu^{4}}{N^{2}}\left(\frac{n_{\nu}^{2}-1}{n_{\nu}^{2}+2}\right)^{2}F_{k,i}\text{\ensuremath{\left(\nu\right)}}
\end{equation}

where $\nu$ is the wavenumber in $\mathrm{cm^{-1}}$, $N$ is the
number density in , $n_{\nu}$ is the refractive index of the gas
at the wavenumber $\nu$, and $F_{k,i}\text{\ensuremath{\left(\nu\right)}}$
is the King correction factor. The scattering cross section of the
gas mixture $\sigma_{R}\left(\nu\right)=\sum X_{i}\sigma_{R,i}\left(\nu\right)$
is the weighted average from all major atmospheric constituents. The
refractive indices and King correction factor functions of $\mathrm{N_{2},\, CO,\, CO_{2},\, CH_{4},\,\mbox{and}\, N_{2}O}$
are taken from \citet{sneep_direct_2005}, while the refractive index
for $\mathrm{H_{2}O}$ is taken from \citet{schiebener_refractive_1990}.

\paragraph{\textit{Clouds}}

Our model accounts for the potential presence of an opaque cloud deck
whose upper surface's altitude is described by a free retrieval parameter.
We assume a wavelength-independent, sharp cutoff of grazing light
beams at the upper end of the cloud deck. The assumption of a sharp
cutoff reasonably captures the effects of typical condensation cloud
layers because, at ultra-violet to near-infrared wavelengths, such
cloud layers become opaque on length-scales that are small compared
to the uncertainty in the radius measurements probed by the transit
observation. The motivation behind modeling the clouds as a sharp
cutoff of grazing light beams is to obtain a zeroth order model capturing
the trends of clouds on the transmission spectrum while using only
one free parameter for cloudsin the retrieval.

\subsubsection{Temperature-Pressure Profile \label{sub:Temperature-Pressure-Profile}}

We use the analytical description for irradiated planetary atmospheres
by \citet{guillot_radiative_2010} with convective adjustments to
approximate a temperature profile that is self-consistent with the
atmospheric opacities and Bond albedo of each model atmosphere. The
motivation behind this gray-atmosphere approach is that (1) its computational
efficiency allows us to obtain temperature-pressure profiles consistent
with the molecular composition for a large number of model atmospheres
and (2) the uncertainties in the atmospheric temperature are dominated
by the uncertainties in the albedo rather than model errors. 

The \citet{guillot_radiative_2010} model describes the horizontally-averaged
temperature profile $\overline{T}$ as a function of optical depth,
$\tau,$ by

\newcommand{\parenthnewln}{\right.\\&\left.}

\begin{align}\begin{split}\label{eq:Guillot}
	&\overline{T^{4}}=\frac{3T_{\mathrm{int}}^{4}}{4}\left\{ \frac{2}{3}+\tau\right\} +\frac{3T_{\mathrm{eq}}^{4}}{4}\left\{ \frac{2}{3}   +\frac{2}{3\gamma} \parenthnewln \left[1+\left(\frac{\gamma\tau}{2}-1\right)e^{-\gamma\tau}\right]+\frac{2\gamma}{3}\left(1-\frac{\tau^{2}}{2}\right)E_{2}\left(\gamma\tau\right)\right\} , \end{split}\end{align}

where $T_{\mathrm{eq}}$ is the planet's equilibrium temperature,
$T_{\mathrm{int}}$ parameterizes the internal luminosity of the planet
(set to 0 in this work), and $\gamma$ is the ratio of the mean visible
and thermal opacity and therefore parameterizes the deposition of
stellar radiation in the atmosphere. We determine the mean opacities
at visible and thermal wavelengths by averaging the line-by-line opacities
weighted by the black body intensity at the effective star temperature
and at the planet's equilibrium temperature, respectively. 

Given a composition of the model atmosphere and the planetary albedo,
we iteratively determine a solution that is self-consistent with the
molecular opacities and in agreement with radiative and hydrostatic
equilibrium. In the process, we check for the onset of convective
instabilities ( $-\frac{dT}{dz}>\Gamma=\frac{g}{C_{p}}$) delimiting
the transition to the convective layer. For the specific heat capacity
$C_{p}$, we assume that the molecular constituents of the atmosphere
are ideal gases. In the convective regime, we adopt the adiabatic
temperature profile. Our requirement to run a large number of models
currently does not allow us to explicitly account for scattering and
re-radiation of a solid surface or clouds in the calculation of the
temperature profile.

\subsubsection{Transmission Model}

The atmospheric transmission spectrum of an extrasolar planet can
be observed when the planet passes in front of its host star. During
this transit event, some of the star's light passes through the optically
thin part of the atmosphere, leading to excess absorption at the wavelength
at which molecular absorption or scattering is strong. We model the
transmission spectrum following the geometry described by \citet{brown_transmission_2001}.
Given the planetary radius parameter,$R_{P,10}$, and the surface
pressure parameter (Section \ref{sub:Parametrization}), we calculate
the radius at the surface. Below this surface radius, we represent
the planet as an opaque disk. Above the surface radius, we calculate
the slant optical depth $\tau(b)$ as a function of the impact parameter
$b$ by integrating the opacity through the planet's atmosphere along
the observer's line of sight. We account for extinction due to molecular
absorption and Rayleigh scattering. Light that is scattered out of
the line of sight is assumed not to arrive at the observer. We then
integrate over the entire annulus of the atmosphere to determine the
total absorption of stellar flux as a function of wavelength. To assess
the fit between the observations and the model spectrum for a given
set of input retrieval parameters, we integrate the transmission spectrum
over the response functions of the individual instrument channels
used in the observations. These simulated instrument outputs serve
in the MCMC method to evaluate the jump probability as described in
the next section.

\subsection{Atmospheric Retrieval\label{sub:Atmospheric-Retrieval}}

\subsubsection{Bayesian Analysis}

We employ the Bayesian framework using the Markov Chain Monte Carlo
(MCMC) technique to calculate the posterior probability density distribution,
$p\text{\ensuremath{\left(\boldsymbol{x}|\boldsymbol{d}\right)}, }$of
the atmospheric parameters, $\boldsymbol{x}$, given the measured
transit depths in each of the instrumental channels, $\boldsymbol{d}$.
According to Bayes' theorem, the posterior distribution is 
\begin{equation}
p\text{\ensuremath{\left(\boldsymbol{x}|\boldsymbol{d}\right)}}=\frac{p\text{\ensuremath{\left(\boldsymbol{x}\right)}}p\text{\ensuremath{\left(\boldsymbol{d}|\boldsymbol{x}\right)}}}{\int p\text{\ensuremath{\left(\boldsymbol{d}|\boldsymbol{x}\right)p\text{\ensuremath{\left(\boldsymbol{x}\right)d\boldsymbol{x}}}}}},\label{eq:Bayesian}
\end{equation}

where $p\left(\boldsymbol{x}\right)$ represents the prior knowledge
or ignorance of the atmospheric parameters. For extrasolar super-Earths,
we currently have little or no prior knowledge of the atmosphere and
therefore aim for an appropriate non-informative prior distribution
(Sections \ref{sub:Prior-probability-distribution} and \ref{par:Compositional-Analysis}).
The term $p\text{\ensuremath{\left(\boldsymbol{d}|\boldsymbol{x}\right)}}$
represents the probability of measuring the transit depths, $\boldsymbol{d}$,
given that the atmospheric parameters are $\boldsymbol{x}$. It is
modeled with the atmospheric \textquotedbl{}forward\textquotedbl{}
model (Section \ref{sub:Atmosphere-"Forward"-Model}) and an estimate
of the uncertainty in the observed transit depths.

\subsubsection{Markov Chain Monte Carlo (MCMC)}

The MCMC technique using the Metropolis-Hastings algorithm offers
an efficient method of performing the integration necessary for the
Bayesian analysis in Equation (\ref{eq:Bayesian}) \citep{gelman_bayesian_2003}.
It has been applied to several other astronomical data sets and problems
\citep[e.g.,][and references therein]{ford_quantifying_2005}. We
use the MCMC technique to determine the best estimates and Bayesian
credible regions for the atmospheric parameters by computing the joint
posterior probability distribution of the atmospheric parameters.
The uncertainty of individual parameters introduced by complicated,
non-Gaussian correlations with other parameters is accounted for in
a straightforward way by marginalizing the joint posterior distribution
over all remaining parameters. 

The goal of our MCMC simulation is to generate a chain of states,
i.e., a chain of sets of atmospheric parameters $\boldsymbol{x}_{i}$,
that are sampled from the desired posterior probability distribution
$p\text{\ensuremath{\left(\boldsymbol{x}|\boldsymbol{d}\right)}}$.
Using the Metropolis-Hastings algorithm, such a chain can be computed
by specifying an initial set of parameter values, $\boldsymbol{x}_{0}$,
and a proposal distribution, $p\text{\ensuremath{\left(\boldsymbol{x}'|\boldsymbol{x}_{n}\right)}}$.
At each iteration, a new proposal state $\boldsymbol{x}'$ is generated
and the fit between the transit observations and the model transmission
spectrum for the proposed set of atmospheric parameters is computed.
The new proposal state $\boldsymbol{x}'$ is then randomly accepted
or rejected with a probability that depends on (1) the difference
between the $\chi^{2}$-fits of the previous state and the proposal
state and (2) the difference in the prior probability between the
previous state and the prior state. A proposal state that leads to
an improvement in the $\chi^{2}$-fit and a higher prior probability
compared to the previous state is always accepted. A proposal state
that leads to a worse fit or a lower prior probability is accepted
according to the jump probability
\begin{equation}
p\text{\ensuremath{\left(\boldsymbol{x}_{n+1}=\boldsymbol{x}'\right)=\exp\left\{ -\frac{1}{2}\left[\chi^{2}\left(\boldsymbol{x}'\right)-\chi^{2}\left(\boldsymbol{x}_{n}\right)\right]\right\} \cdot\frac{p\text{\ensuremath{\left(\boldsymbol{x}'\right)}}}{p\text{\ensuremath{\left(\boldsymbol{x}_{n}\right)}}}}},
\end{equation}

where we assumed Gaussian uncertainty in the observations and
\begin{equation}
\chi^{2}=\sum_{k=1}^{n_{\mathrm{obs}}}\frac{\left(D_{k,\mathrm{model}}-D_{k,\mathrm{obs}}\right)^{2}}{\sigma_{k}^{2}}
\end{equation}

is the measure of fit between the observed transit depths, $D_{k,\mathrm{obs}}$,
and the model output, $D_{k,\mathrm{model}}$. The probabilities $p\text{\ensuremath{\left(\boldsymbol{x}_{n}\right)}}$and
$p\text{\ensuremath{\left(\boldsymbol{x}'\right)}}$are the prior
probabilities of the previous and the proposal state. If the proposal
state is rejected, the previous state will be repeated in the chain.

\subsubsection{Parallel Tempering}

A simple Metropolis-Hastings MCMC algorithm can fail to fully explore
the target probability distribution, especially if the distribution
is multi-modal with widely separated peaks. The algorithm can get
trapped in a local mode and miss other regions of the parameter space
that contain significant probability. The trapping problem is expected
for atmospheric retrieval of extrasolar planets for which only very
sparse data are available. The challenge faced is similar to the one
encountered in finding the global minimum of a nonlinear function.

We address the challenge of a potentially multi-modal probability
distribution by adopting a parallel tempering scheme \citep{gregory_bayesian_2005-1}
for our atmospheric retrieval method. In parallel tempering, multiple
copies of the MCMC simulation are run in parallel, each using a different
\textquotedbl{}temperature\textquotedbl{} parameter, $\beta$. The
tempering distributions are described by

\begin{equation}
p\text{\ensuremath{\left(\boldsymbol{x}|\boldsymbol{d},\beta\right)}}=p\text{\ensuremath{\left(\boldsymbol{x}\right)}}p\text{\ensuremath{\left(\boldsymbol{d}|\boldsymbol{x}\right)^{\beta}}}
\end{equation}

One of the simulated distributions, the one for which we choose $\beta=1,$
is the desired target distribution. The other simulations correspond
to a ladder of distributions at higher temperature with $\beta$ ranging
between 0 and 1. For $\beta\ll1$, the simulated distribution is much
flatter and a wide range of the parameter space is explored. Random
swaps of the parameter states between adjacent simulations in the
ladder allow for an exchange of information across the different chains.
In the higher temperature distributions ($\beta\ll1$), radically
new configurations are explored, while lower temperature distributions
($\beta\sim1$) allow for the detailed exploration of new configurations
and local modes.

The final inference on atmospheric parameters is based on samples
drawn from the target probability distribution ($\beta=1$) only.
To probe the convergence, we perform multiple independent parallel-tempering
simulations of the target probability distribution with starting points
dispersed throughout the entire parameter space.

\subsubsection{Ignorance Priors \label{sub:Prior-probability-distribution}}

One challenge in atmospheric retrieval is that even for the most simple
atmospheric parameterizations, some parameters describing the composition
and state of the atmosphere might not be constrained well by the observations.
In this regime, it is important to choose an appropriate, non-informative
prior probability distribution on the parameters. One advantage of
the Bayesian approach over traditional frequentist approaches is that
we can explicitly state our choice of the prior probability distribution.
Many approaches, e.g., constant-$\Delta\chi^{2}$ boundaries, usually
implicitly assume a uniform prior. While in many cases the uniform
prior seems like a reasonable choice, it is worth noting that the
uniform prior is variant under reparameterization. For example, a
uniform prior for $\log\left(x\right)$ will not be a uniform prior
for $x$ \citep{gelman_bayesian_2003}, therefore the obtained results
can depend on the choice of parameterization.

In this work, we use a uniform prior on the radius ratio parameter,
$\left(R_{p}/R_{*}\right)_{10}$, and the planetary albedo, $A$.
The surface pressure, $P_{\mathrm{surf}}$, is a \textquotedbl{}scale
parameter\textquotedbl{} for which we do not know the order of magnitude
a priori. We therefore choose a Jeffrey prior for the surface pressure,
i.e., a uniform prior for $\log\text{\ensuremath{\left(P_{\mathrm{surf}}\right)}}$.
Since an infinite surface pressure may agree with the observational
data in the same way a sufficiently high finite value does, the posterior
distribution can remain unnormalizable unless a normalizable prior
distribution is chosen. To ensure a normalizable posterior, we set
an upper bound on the prior at $p=100\,\mathrm{bar}$ . Higher surface
pressures are not considered because atmospheres of plausible compositions
will be optically thick to the grazing star light at higher pressure
levels. 

The mixing ratios of the molecular gases are also scale parameters,
suggesting that the usage of a Jeffrey prior for each of the mixing
ratios would be appropriate. The constraint that the mixing ratios
must add up to unity, however, prevents the assignment of a Jeffrey
prior for the individual mixing ratio. We therefore introduce a reparameterization
as discussed in the following section.

\subsubsection{Centered-Log-Ratio Transformation for Mixing Ratios of Atmospheric
Constituents\label{par:Compositional-Analysis}}

Since the mixing ratios of the molecular species in the atmosphere
present parts from a whole, they must satisfy the constraints
\begin{equation}
0<X_{i}<1\,\textrm{and}\label{eq:}
\end{equation}

\begin{equation}
\sum_{i=1}^{n}X_{i}=1,\label{eq:-1}
\end{equation}

where $n$ is the number of gases in the atmosphere. For the statistical
analysis of the mixing ratios, it is important to recognize that the
sample space of a composition is \textit{not} the full Euclidean space
$\mathbb{R}^{n}$, for which most statistical tools were developed,
but only the restricted ($n-1$)-dimensional space formally known
as the simplex of $n$ parts, $\mathbb{S}^{n}$. The simplex includes
only sets of mixing ratios for which the components sum up to 1. As
a result, the total number of free parameters describing the molecular
composition is reduced by one. The mixing ratio of the $n$th species
$X_{n}$ can be calculated directly from the mixing ratios $X_{1}...X_{n-1}$.

In this subsection, we present a reparameterization for the mixing
ratios that allows for efficient sampling of the full simplex with
MCMC, while ensuring that all $n$ molecular species may range across
the complete detectable range, e.g., $10^{^{-12}}<X_{i}<1$, with
a non-zero prior probability, and that the results are permutation
invariant, i.e., independent of which molecule was chosen to be the
$n$th species.

\begin{figure}[tb]
\includegraphics[clip,width=1\columnwidth]{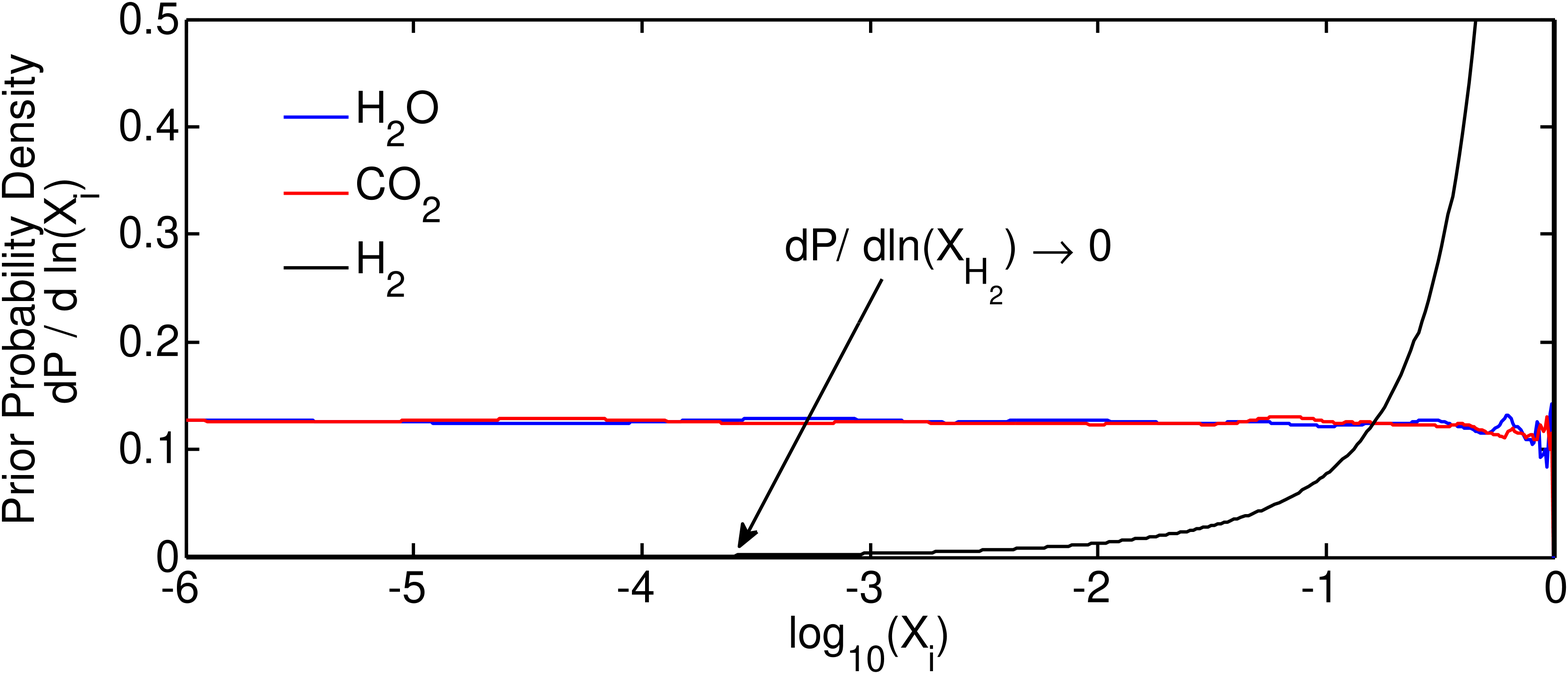}

\includegraphics[clip,width=1\columnwidth]{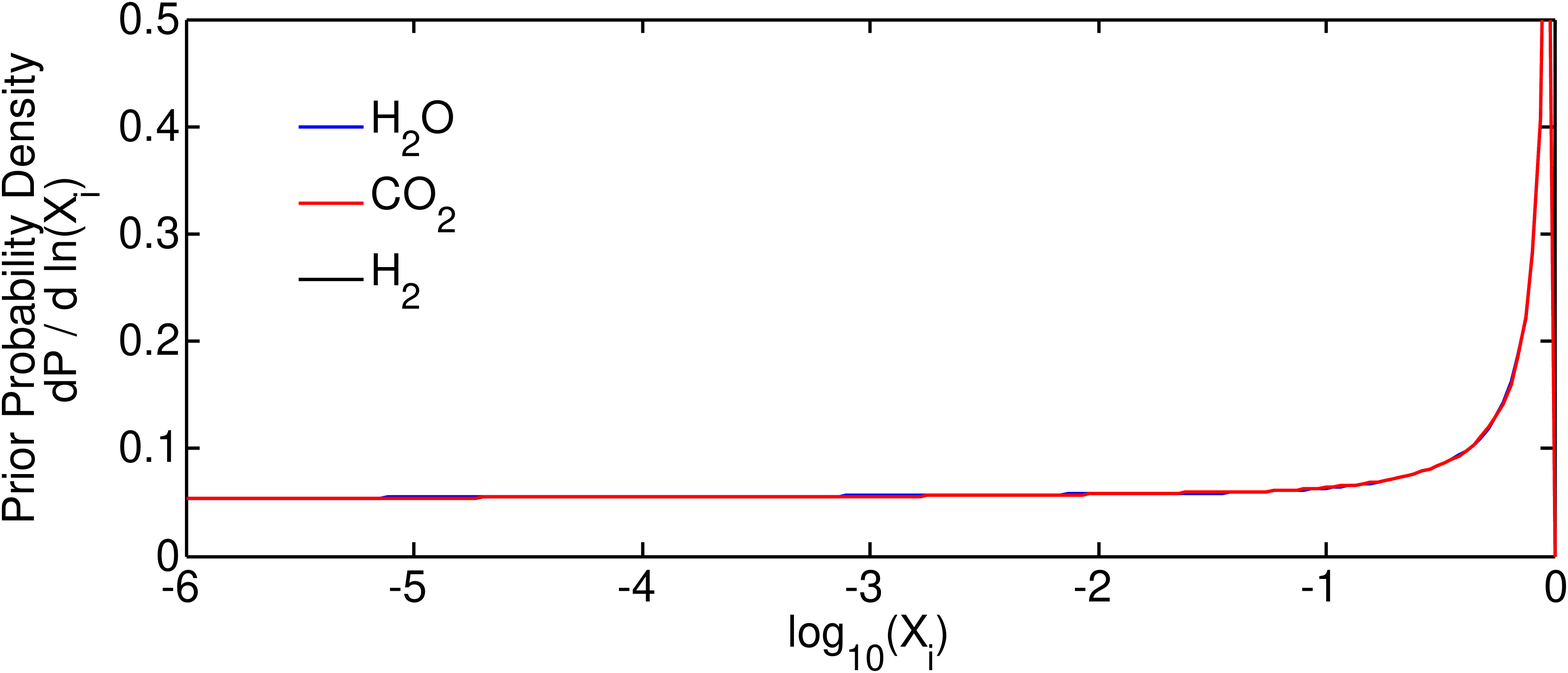}

\noindent \centering{}\caption{Marginalized prior probability distribution for the mixing ratios
in a mixture of three gases. The upper panel illustrates the prior
probabilities for a parameterization in which the gases $\mbox{H}_{2}\mbox{O}$
and $\mbox{CO}_{2}$ are described by free parameters and $\mbox{H}_{2}$
is set to fill the remainder of the atmosphere. Assigning a Jeffrey
prior, i.e., $dP/d\ln\left(X_{i}\right)=\mathrm{const}$ to all gases
except one leads to a description that is permutation variant and
strongly favors compositions with a high abundance of the remaining
gas. Compositions with low amount of $\mbox{H}_{2}$ are excluded
by the prior because the prior probability rapidly approaches zero
for $X_{\mbox{H}_{2}}<1\%$ . The bottom panel shows the prior probabilities
for the mixing ratios using the center-log-ratio transformation introduced
in this work. The prior probability for all gases in the mixture is
identical, thereby providing permutation-invariant results. The prior
probability distribution of \textit{all} gases approaches the Jeffrey
prior at mixing ratios below $\thicksim$20\%, and is, therefore,
highly favorable for scale parameters. The divergence to infinity
is only of theoretical nature. Once the signature of one gas is detected
in the spectrum, the posterior probability of all other gases at $X_{i}=100\%$
will go to zero. \label{fig:Marginalized-prior-probability}}
\end{figure}

Previous work on atmospheric retrieval implicitly accounted for the
constraints in Equations (\ref{eq:}) and (\ref{eq:-1}) by using
free parameters only for mixing ratios of the minor atmospheric gases
and assuming that the remainder of the atmosphere is filled with the
a-priori known main constituent. This approach is feasible for the
inference of gas mixing ratios in hot Jupiters and Solar System planets
because the main constituents of the atmospheres, i.e., $\mbox{H}_{2}$,
are known a priori. In a Bayesian retrieval approach, in which we
do not know the main constituent of the atmosphere, however, parameterizing
the abundances of minor species and assuming a main species is unfavorable.
Assigning a Jeffrey prior, i.e. a uniform prior on the logarithmic
scale, for $n-1$ mixing ratios leads to a highly asymmetric prior
that favors a high abundance of the $n$th species (Figure \ref{fig:Marginalized-prior-probability}
(top)).  In addition, in cases with a low abundance of the $n$th
species, we find that the asymmetric parameterization leads to serious
convergence problems in the numerical posterior simulation with MCMC.

To circumvent the drawbacks of highly asymmetric priors in the interpretation
of the results as well as the numerical convergence problems due a
highly asymmetric parameterization, we use the centered-log-ratio
transformation to reparameterize the composition \citep{aitchison_statistical_1986,pawlowsky-glahn_compositional_2006}.
The centered-log-ratio transformation is commonly used in geology
and social sciences for the statistical analysis of compositional
data \citep[e.g.,][]{pawlowsky-glahn_compositional_2006}, we find
that it also enables the MCMC technique to efficiently explore the
posterior distribution of the atmospheric composition across the complete
simplicial sample space.

For a mixture of $n$ gases, the centered-log-ratio transformation
of the $i$-th molecular species is defined as

\begin{equation}
\xi_{i}=\mathrm{clr}\left(X_{i}\right)=\ln\frac{X_{i}}{g(\boldsymbol{x})},
\end{equation}

where 

\begin{equation}
g(\boldsymbol{x})=\left(\prod_{j=1}^{n}X_{i}\right)^{1/n}=\exp\left(\frac{1}{n}\sum_{j=1}^{n}\ln X_{i}\right)
\end{equation}

is the geometric mean of all mixing ratios $X_{1}...X_{n}$.

\begin{figure}[tb]
\noindent \begin{centering}
\includegraphics[clip,width=0.7\columnwidth]{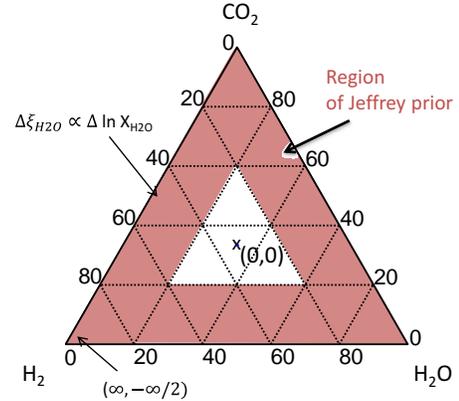}
\par\end{centering}

\noindent \centering{}\caption{Simplicial sample space for a mixture of three gases illustrated in
a ternary diagram. Using the centered-log-ratio transformation for
the mixing ratios of the atmospheric gases, we obtain a symmetric
parameterization of the composition in which all molecular species
are treated equally, while simultaneously ensuring that the sum of
the mixing ratios is unity. The zero point of the transformed mixing
ratios, $\xi_{i}$, is at the center of the simplex. Toward the edges
of the sample space, i.e., for low mixing ratios of one or more gases,
the differences in the transformed mixing ratio, $\xi_{i}$, scale
with $\ln\left(X_{i}\right)$. The scaling provides a region in which
the prior probability $dP/d\ln(X_{i})$ remains constant (red) and
allows the MCMC to efficiently sample down to exponentially small
mixing ratios for all molecular species. \label{fig:Prior-probability-density-1}}
\end{figure}

Each of the compositional parameters $\xi_{i}$ may range between
$-\infty$ and $+\infty$, where the limit $\xi_{i}\rightarrow-\infty$
indicates that $i$th species is of extremely low abundance with respect
to the other molecular species, while $\xi_{i}\rightarrow+\infty$
indicates that the $i$-th species is abundant in the atmosphere.
The composition $\boldsymbol{\xi}=\left[0,0,\ldots,0\right]$describes
the center of the simplex at which all molecular species are equally
abundant (Figure \ref{fig:Prior-probability-density-1}). The only
constraints on the transformed compositional parameters is $\sum_{i=1}^{n}\xi_{i}=0$.

A fully permutation-invariant description is obtained by using the
centered-log-ratio transformed mixing ratios $\xi_{1}\ldots\xi_{n-1}$
as the free parameters and assigning a uniform prior for all vectors
in the $\mathbf{\xi}$-space for which $X_{i}>10^{-12}$ for all $i=1\ldots n$.
As the distances in the sample space spanned by $\xi_{1}\ldots\xi_{n-1}$
scale with the differences in $\ln X_{i}$ for small mixing ratios,
the MCMC can efficiently sample the complete space, even if the mixing
ratios vary over several orders of magnitude.When transformed back
into the Euclidean space of the mixing ratios, $X_{i}$, we obtain
prior probabilities for each of the mixing ratios that have the properties
of a Jeffrey prior below $X_{i}\lesssim20\%$ (Figure \ref{fig:Prior-probability-density-1}).
The properties of a Jeffrey prior are highly favorable for scale parameters
such as the mixing ratios. The increase in the prior toward mixing
ratios $\gtrsim0.1$ is a direct consequence of the fact that one
or more molecular species in the atmosphere inevitably needs to have
a significant abundance. The divergence toward infinity at $\log X_{i}$
is of no practical relevance. If a single molecular species is detectable
in the spectrum, then mixing ratios of 100\% are excluded for all
other species by the data .

\subsection{Inputs}

The primary inputs to the retrieval method presented here are spectral
and/or photometric observations of the wavelength dependent transit
depths during the primary transit, $\left(R_{p}/R_{*}\right)^{2}$.
Accurate estimates of the observational error bars are of particular
importance because they can significantly affect the constraints and
conclusions made from the observations. Ideally, the spectral data
would not be binned to reduce the apparent error bars. Binning of
spectral data always leads to a loss of information and should only
be done if is required to compensate for systematics. 

The spectra from the primary transit can be augmented by secondary
eclipse measurements constraining the planetary albedo (or the atmospheric
temperature) by including the information in the prior probability
distribution. An improved estimate of the temperature or planetary
albedo can lead to improved constraints in all composition parameters.
If no such observations are available, the retrieved uncertainties
in the composition will fully account for the uncertainty in the planetary
albedo.

\subsection{Outputs\label{sub:Output}}

The output of the atmospheric retrieval is the posterior probability
density distribution, $p\text{\ensuremath{\left(\boldsymbol{x}|\boldsymbol{d}\right)}}$,
of the retrieval parameters discussed in Section \ref{sub:Parametrization}.
This multidimensional distribution encodes our complete state of knowledge
of the atmospheric parameters in the light of the available observations..
To illustrate our state of knowledge of a single parameter, we marginalize
the posterior density distribution over all remaining parameters.
For well-constrained parameters, one can summarize our knowledge of
the parameter in just a few numbers, i.e., the most likely estimates
and a set of error-bars and correlation coefficients. Depending on
the nature of the observational data, however, we may obtain a posterior
distribution that is not well-described by single best estimate plus
the uncertainty around this estimate. For example, a multi-modal distribution
would be indicative of multiple possible solutions. Highly asymmetric
posterior distributions or only one-sided bounds will be obtained
if the observations constrain the parameter only in one direction.

We can also compute the constraints on atmospheric properties that
do not serve as free parameters in our retrieval methods, such as
the mean molecular mass, the total atmospheric mass above the surface,
the mixing ratios by mass, or the elemental abundances. A set of the
retrieval parameters introduced in Section \ref{sub:Parametrization}
entirely describes the state of well-mixed, one-dimensional atmospheres.
For each set of retrieval parameters in the chain obtained from the
MCMC simulations, we can, therefore, compute the any other atmospheric
property from the retrieval parameters. In this way, we obtain a new
chain for the desired atmospheric property that, interpreted as a
sample from the marginalized distribution of the atmospheric property,
can be used to infer constraints on the atmospheric properties by
comparing the distribution to the equivalently obtained prior distribution.

In this work, we present constraints on the mean molecular mass, total
atmospheric mass above the cloud-deck/surface, and relative elemental
abundances (Equations 10-12),

\begin{equation}
\mu_{\mathrm{mix}}=\sum_{i=1}^{n}\mu_{i}X_{i}
\end{equation}

\begin{equation}
M_{\mathrm{atm}}=\int4\pi r^{2}\rho\left(r\right)dr
\end{equation}

\begin{equation}
q_{j}=\frac{\sum_{i}^{n_{\mathrm{mol}}}X_{i}n_{i,j}}{\sum_{j}^{n_{\mathrm{elem}}}\sum_{i}^{n_{\mathrm{mol}}}X_{i}n_{i,j}}
\end{equation}

where $q_{j}$ is the relative abundance of the elemental species
$j$, $X_{i}$ is the mixing ratio of molecule $i$, and $n_{i,j}$
is the number of atoms of elemental species $j$ in molecule $i$.

\section{Synthetic Observations of Super-Earth Transmission Spectra \label{sec:Synthetic-Observations-of}}

The goal of the quantitative analysis presented in this work is to
explore which constraints on the atmospheric properties of super-Earth
exoplanets can be extracted from low-noise transmission spectra in
the coming decades. In this section, we describe synthetic, low-noise
observations of the transmission spectra of three different, hypothetical
types of hot super-Earths transiting nearby M-stars as they may be
obtained with the \textit{James Webb Space Telescope} (\textit{JWST}).
To make the results most relevant in the context of current observational
efforts, we adopt the stellar, orbital, and planetary parameters of
the super-Earth GJ~1214b \citep{charbonneau_super-earth_2009}.

\subsection{Atmospheric Scenarios\label{sub:Planetary-Scenarios}}

\paragraph{``\textit{Hot Halley world}.''}

The first scenario we consider is a volatile-rich super-Earth \citep{kuchner_volatile-rich_2003,leger_new_2004}.
The motivation for this scenario is to investigate the retrieval results
for an atmosphere that is predominately composed of absorbing gases.
For our specific case, we consider a scenario in which the planet
has accreted ices with the elemental abundances of the ices identical
to those in the Halley comet in the Solar System \citep{jessberger_chemical_1991}.
Some of the ices may have evaporated at the high equilibrium temperature
and formed an atmosphere around the planet. We assume a well-mixed
atmosphere around the planet whose chemical composition is calculated
from chemical equilibrium at the 1 bar level. The resulting atmosphere
is composed of $\mathrm{H_{2}O}$ (69.5\%), $\mathrm{CO_{2}}$ (13.9\%),
$\mathrm{H_{2}}$ (11.8\%), $\mathrm{CH_{4}}$ (2.6\%), and $\mathrm{N_{2}}$
(2.2\%). All mixing ratios are givenas volume mixing ratios. The atmosphere
is assumed to be clear and sufficiently thick such that no surface
affects the transmission spectrum in this scenario. (Table \ref{tab:Molecular-abundances-of}).

\paragraph{``\textit{Hot nitrogen-rich world}.''}

For the second scenario we consider a nitrogen-dominated atmosphere
representative of a rocky planet with an outgassed atmosphere similar
to the atmospheres of Earth and Titan. The motivation for this scenario
is to investigate the retrieval results for an atmosphere that is
predominately composed of a spectrally-inactive gas that has no directly
observable features in the spectrum. We chose an atmosphere dominated
by $\mathrm{N_{2}}$ (95.4\%) and rich in $\mathrm{CH_{4}}$ (3.5\%),
$\mathrm{CO_{2}}$ (1\%) and $\mathrm{H_{2}O}$ (0.1\% ) with a rocky
surface at 1 bar.

\paragraph{``\textit{Hot mini-neptune.}''}

The third scenario is a super-Earth with a thick hydrogen/helium envelope
that has experienced a formation history similar to those of the giant
planets. While we assume a primordial atmosphere, we deliberately
chose an example that is away from solar abundance and chemical equilibrium.
The motivation for this scenario is to demonstrate the retrieval for
a scenario that does not correspond to our preconceived ideas. The
atmosphere we consider is composed of 84.9\% $\mathrm{H_{2}}$, 13.1\%
$\mathrm{He}$, and 2\% $\mathrm{H_{2}0}$, and has small mixing ratios
of $\mathrm{CO_{2}}$ ($10^{-6})$ and $\mathrm{CH_{4}}$ ($10^{-7})$.
We consider the presence of an opaque cloud deck of unknown nature
at the 100 mbar level.

\begin{table*}[tb]
\begin{centering}
\begin{tabular}{lccccccc}
\noalign{\vskip\doublerulesep}
Planet Scenario & $X_{\mathrm{H_{2}O}}$ & $X_{\mathrm{CO_{2}}}$ & $X_{\mathrm{CH_{4}}}$ & $X_{\mathrm{H_{2}}}$ & $X_{\mathrm{He}}$ & $X_{\mathrm{N_{2}}}$ & Surface\tabularnewline[\doublerulesep]
\hline 
\noalign{\vskip\doublerulesep}
\noalign{\vskip\doublerulesep}
Hot Halley world & 69.5\% & 13.9\% & 2.6\% & 11.8\% & $\approx0$ & 2.2\% & None\tabularnewline[\doublerulesep]
\noalign{\vskip\doublerulesep}
\noalign{\vskip\doublerulesep}
Hot nitrogen-rich world & 0.1\% & 1\% & 3.5\% & $\approx0$ & $\approx0$ & 95.4\% & Rocky surface at$1\,\mathrm{bar}$\tabularnewline[\doublerulesep]
\noalign{\vskip\doublerulesep}
\noalign{\vskip\doublerulesep}
Hot mini-neptune & 2\% & $10^{-6}$ & $10^{-7}$ & 84.9\% & 13.1\% & $\approx0$ & Cloud deck at $100\,\mathrm{mbar}$\tabularnewline[\doublerulesep]
\noalign{\vskip\doublerulesep}
\end{tabular}
\par\end{centering}

\caption{Mixing Ratios Of Molecular Constituents and Surface Pressure for the
Three Super-Earth Scenarios Used to Generate Synthetic Transmission
Spectra. \label{tab:Molecular-abundances-of}}
\end{table*}

\subsection{Observation Scenarios \label{sub:Observation-Scenarios}}

For each of the three atmospheric scenarios, we simulate high-resolution
transmission spectra ($R>10^{5})$ and model the output of the \textit{JWST}
NIRSpec instrument covering the spectral range between $0.6$ and
$5\,\mathrm{\mu m}$. We assume that the transit depths in the individual
channels of \textit{JWST} NIRSpec can be determined to within 20\%
of the shot noise limit. To compute the photon flux for each spectral
channel individually, we scale the spectrum of a typical M4.5V star
\citep{segura_biosignatures_2005} to the apparent brightness of GJ~1214.
For \textit{JWST}, we adopt an effective diameter of the primary mirror
of $6.5\,\mathrm{m}$ and a throughput before the instrument of 0.88
\citep{deming_discovery_2009}. We consider the spectral resolution
for observations using the $R$=100 $\mbox{CaF}_{2}$ prism on NIRSpec
($R=30\ldots280$). Our noise model adopts a total optical transmission
for the NIRSpec optics after the slit of 0.4 and a quantum efficiency
for the HgCdTe detector of 0.8 \citep{deming_discovery_2009}. We
do not include any slit losses because the large aperture of \textit{JWST}
will encompass virtually all of the energy in the point source function.
We find that read noise ($6\, e^{-}$per Fowler 8) and dark current
($0.03\,\mathrm{e^{-}\, s^{-1}}$) are insignificant compared to photon
noise. We account for a $\sim20\%$ loss of integration time due to
the resetting and reading-out of the detector, based on the expected
saturation time of 0.43~sec for the brightest pixels on the NIRSpec
detector for GJ~1214 (de Wit, personal communication). For a first
order estimate of the observational errors, we neglect the wavelength
dependence of the grating blaze function.

Given the instrument properties, we calculate the expected variances
of the in-transit and out-of-transit fluxes due to shot noise and
calculate the expected error in the observed transit depth. We assume
that the total observation time used to measure the baseline flux
before and after the transit equals the transit duration. We stack
10 synthetic transit observations for the high mean molecular mass
atmospheres of the \textquotedbl{}hot Halley world\textquotedbl{}
and \textquotedbl{}hot nitrogen-rich world\textquotedbl{} scenarios
and use only a single transit observation for the more easily detectable
hydrogen-dominated \textquotedbl{}hot mini-neptune\textquotedbl{}
scenario.

\section{Results}

Our most significant finding is that a \textit{unique} constraint
on the mixing ratios of the absorbing gases and up to two spectrally~inactive
gases is possible with moderate-resolution transmission spectra. Assuming
a well-mixed atmosphere and that $\mathrm{N}_{2}$ and a primordial
mix of $\mathrm{H_{2}+He}$ are the only significant spectrally~inactive
components, we can fully constrain the molecular composition of the
atmosphere.  We also find, however, that even a robust detection
of a molecular absorption feature (>10$\sigma$) can be insufficient
to determine whether a particular absorber is the main constituent
of the atmosphere ($X_{i}>50\%$) or just a minor species with a mixing
ratio of only a less than 0.1\%, if we do not observe the signature
of gaseous Rayleigh scattering.

In this section, we first conceptually identify the features in the
spectrum that are required to uniquely constrain the compositions
of general exoplanet atmospheres (Section \ref{sub:Atmospheric-Composition}).
Based on the conceptual understanding, we then present numerical results
from the MCMC retrieval analysis for synthetic \textit{JWST} NIRSpec
observations of three scenarios for the super-Earth GJ~1214b (Section
\ref{sub:Quantitative-constraints-on}).

\subsection{\textit{Uniquely} Constraining Exoplanet Atmospheres \label{sub:Atmospheric-Composition}}

Identifying absorbing molecules by their spectral features is conceptually
straightforward, as molecules generally absorb at distinct wavelengths.
Constraining the mixing ratios of the atmospheric gases is more complicated
because the observable transmission spectrum depends not only on the
mixing ratios of the absorbers, but also on the exact planetary radius
(as measured by the radius at the reference pressure level, $R_{P,10}$),
the surface or cloud-top pressure, and the mean molecular mass of
the background atmosphere. The absorber mixing ratios may therefore
remain unconstrained over several orders of magnitude despite strong
detections of a molecular absorption features in the near-infrared
wavelength range. The difficulty in constraining the mixing ratios
of the atmospheric constituents was not discovered in previous work
on atmospheric retrieval because hot Jupiters were assumed to be cloud-free
and the mean molecular mass of their hydrogen-dominated atmospheres
was known a priori \citep[e.g.,][]{madhusudhan_temperature_2009}.
In the following, we explain which observables from different parts
of the spectrum must be combined to successfully constrain the composition
of a \textit{general} exoplanet atmosphere. 

\begin{figure*}[tb]
\begin{centering}
\includegraphics[clip,width=1\textwidth]{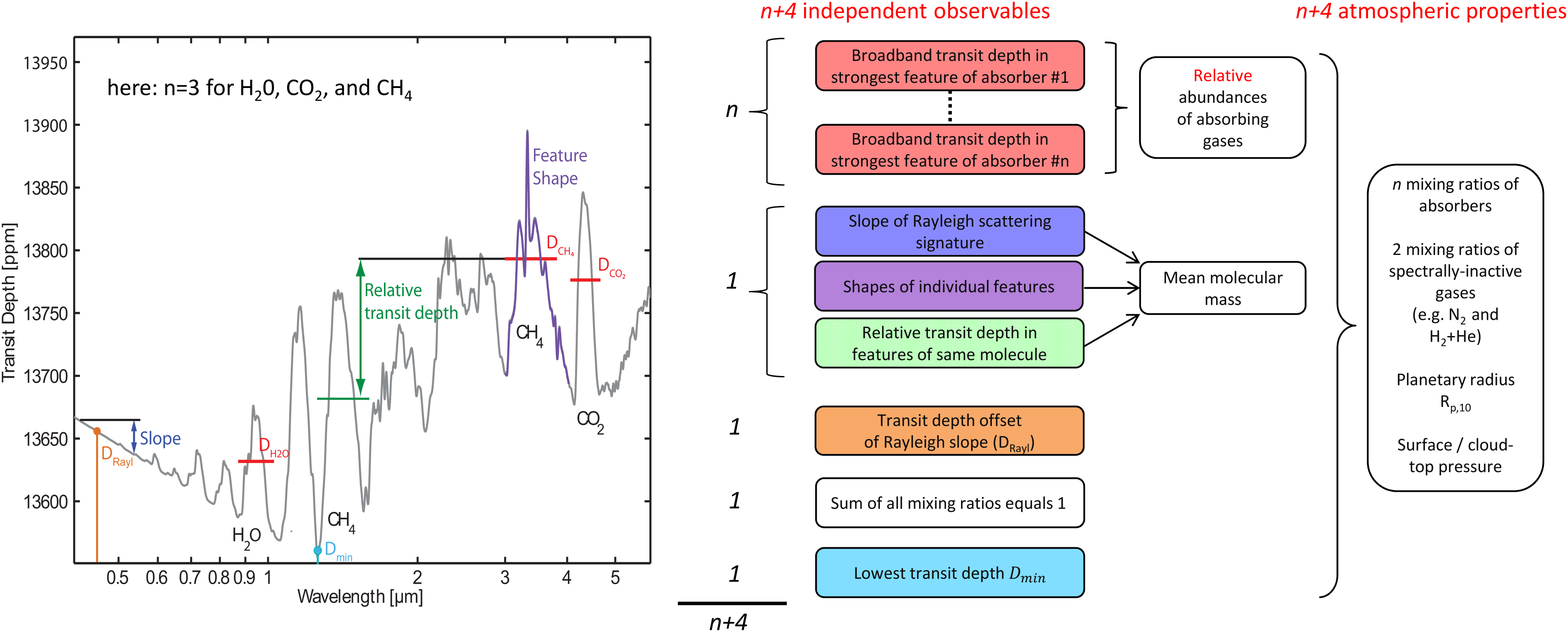}
\par\end{centering}

\caption{Unique constraints on the atmospheric properties based on observables
in the transmission spectrum. The transmission spectrum of an atmosphere
with $n$ relevant absorbers contains $n+4$ independent pieces of
information that constrain the $n$ mixing ratios of these absorbers,
up to two mixing ratios of the two spectrally inactive components
$\mbox{H}_{2}+\mbox{He}$ and $\mbox{N}_{2}$, the planetary radius
at a reference pressure level, $R_{P,10}$, and the surface/cloud-top
pressure. The left panel illustrates conceptually the individual observables
in the transmission spectrum that carry the $n$+4 pieces of information
for an example with $n=3$ absorbers. For well-mixed atmospheres,
the three observables ``Slope of the Rayleigh signature,'', ``Shapes
of individual features,'' and ``Relative transit depths in features
of same molecule'' are redundant and provide only one independent
piece of information. Note that to uniquely constrain \textit{any}
of the $n+4$ atmospheric properties on the far right, \textit{all}
$n+4$ pieces of information need to be available, unless additional
assumptions are made. \label{fig:Flow-Chart}}
\end{figure*}

The transmission spectrum of an atmosphere with $n$ relevant absorbers
provides $n+4$ independent observables (Figure \ref{fig:Flow-Chart}).
Combined, these $n+4$ observables can be used to constrain the $n$
unknown mixing ratios of the absorbing gases, the mixing ratios of
up to two spectrally inactive gases (e.g., $\mbox{N}_{2}$ and primordial
$\mbox{H}_{2}+\mbox{He}$), the planetary radius at reference pressure
level, and the pressure at the surface or upper cloud deck. The remaining
information in the transmission spectrum is highly redundant with
the $n+4$ independent observables. 

The $n+4$ independent observables are as follows. For each of the
$n$ absorbers, the broadband transit depths in the strongest features
provide one independent observable. By measuring and comparing the
broadband transit depths in the absorption features of different molecules,
we can directly determine the \textit{relative} abundances of the
absorbing gases in the atmosphere (Section \ref{sub:Abundance-ratios-of}).
For example, given the broadband transit depths in the 3.3~\textmu{}m
$\mbox{CH}_{4}$ feature and in the 4.3~\textmu{}m $\mbox{CO}_{2}$
feature, we can determine that there must be ``x'' times more $\mbox{CO}_{2}$
than $\mbox{CH}_{4}$ in the atmosphere. If the feature of one molecular
absorber is not present, transit depth measurements at wavelengths
for which the absorption cross sections of the molecular species are
high can still provide an upper limit on the absorber abundance \textit{relative}
to the other absorbers.

Next, we have a total of \textit{one} additional piece of information
from either (1) the linear slope of the Rayleigh signature, (2) the
shapes of individual features, or (3) the relative transit depths
in features of the same molecule. The information from the three observables
is highly redundant.  From one of the three observables, we can directly
constrain the scale height, and, given an approximate estimate of
the atmospheric temperature, we can obtain an estimate of the mean
molecular mass (Section \ref{sub:Mean-molecular-mass}). Importantly,
for general atmospheres that may contain clouds, it is the slope at
which the transit depth changes as a function of the extinction cross
section thatenables us to measure the mean molecular mass. The overall
transit depth variation as currently discussed in many papers on the
super-Earth GJ~1214b \citep[e.g.,][]{miller-ricci_nature_2010,bean_optical_2011,croll_broadband_2011}
measures the mean molecular mass only for cloud-free atmospheres.

\textit{Three} additional independent constraints are provided by
the transit depth offset of the Rayleigh slope, the fact that all
mixing ratios must sum to 1, and the measure of the lowest transit
depths in the spectrum. Comparing the transit depth offset of the
Rayleigh slope and the transit depths at near-infrared wavelengths
provides us with a measure of the amount of spectrally inactive gas
in the atmosphere. Given all previously discussed observables, the
lowest transit depths in the spectrum allow us to independently constrain
the surface/cloud-top pressure (Section \ref{sub:Surface-Pressure}).
If the surface/cloud-top is at a deep layer in the atmosphere and
the molecular opacities across the observed wavelength range are high,
a direct detection of a surface is not possible. In this case, the
minimum transit depth will provide a lower limit on the surface/cloud-top
pressure.

Note that we need \textit{all} $n+4$ observables together in order
to determine \textit{any} of the atmospheric parameters uniquely.
If a single piece of the puzzle is missing, e.g., the transit depths
at short wavelengths are not observed, then the composition, including
the volume mixing ratios of the absorbers, will stay weakly constrained,
even if we have detected the feature with high significance.

\subsubsection{\noindent Relative Abundances of Absorbing Gases\label{sub:Abundance-ratios-of}}

\selectlanguage{british}%
\begin{figure*}[tb]
\hfill{}\foreignlanguage{english}{\includegraphics[width=0.5\textwidth]{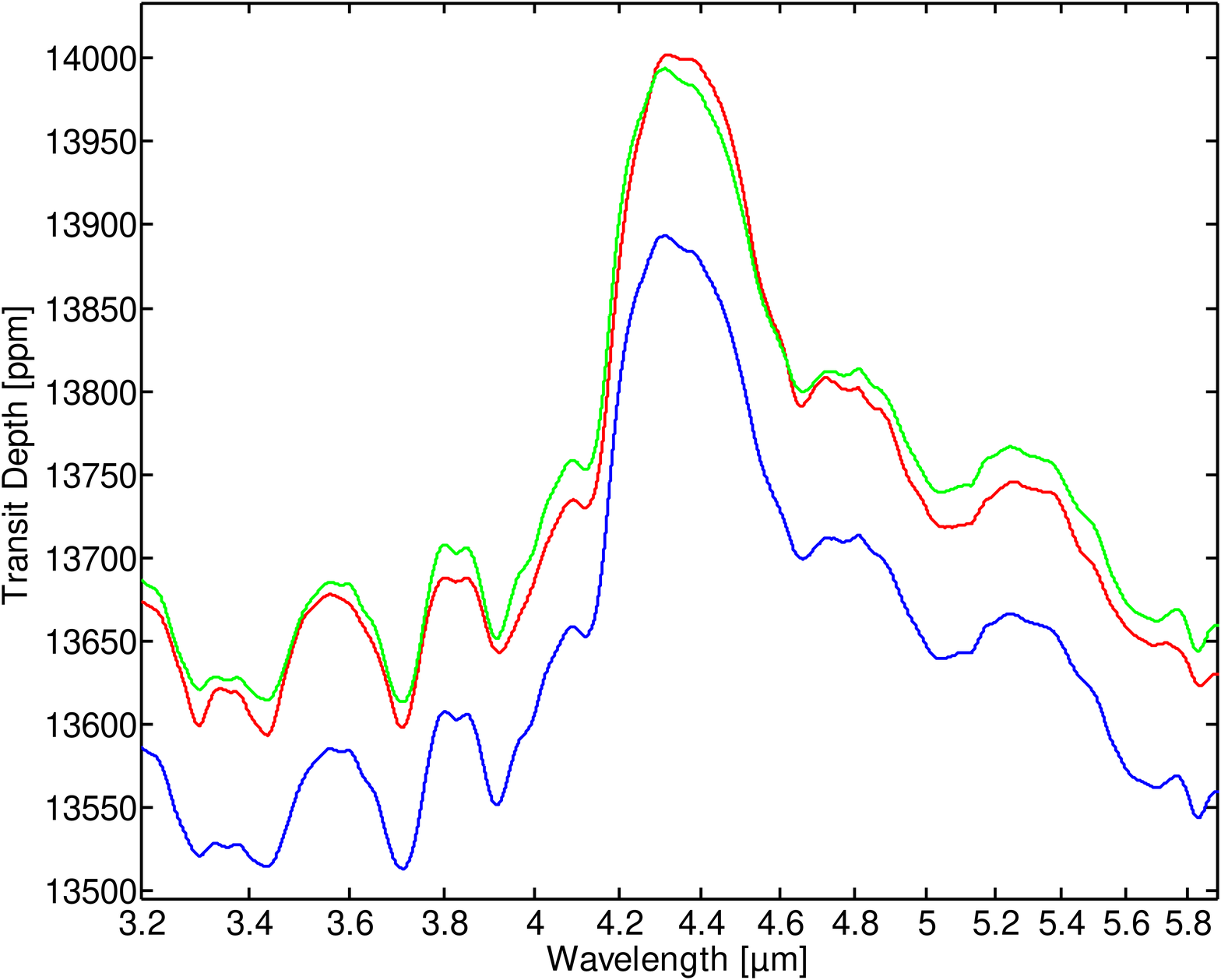}}\hfill{}\foreignlanguage{english}{\includegraphics[width=0.5\textwidth]{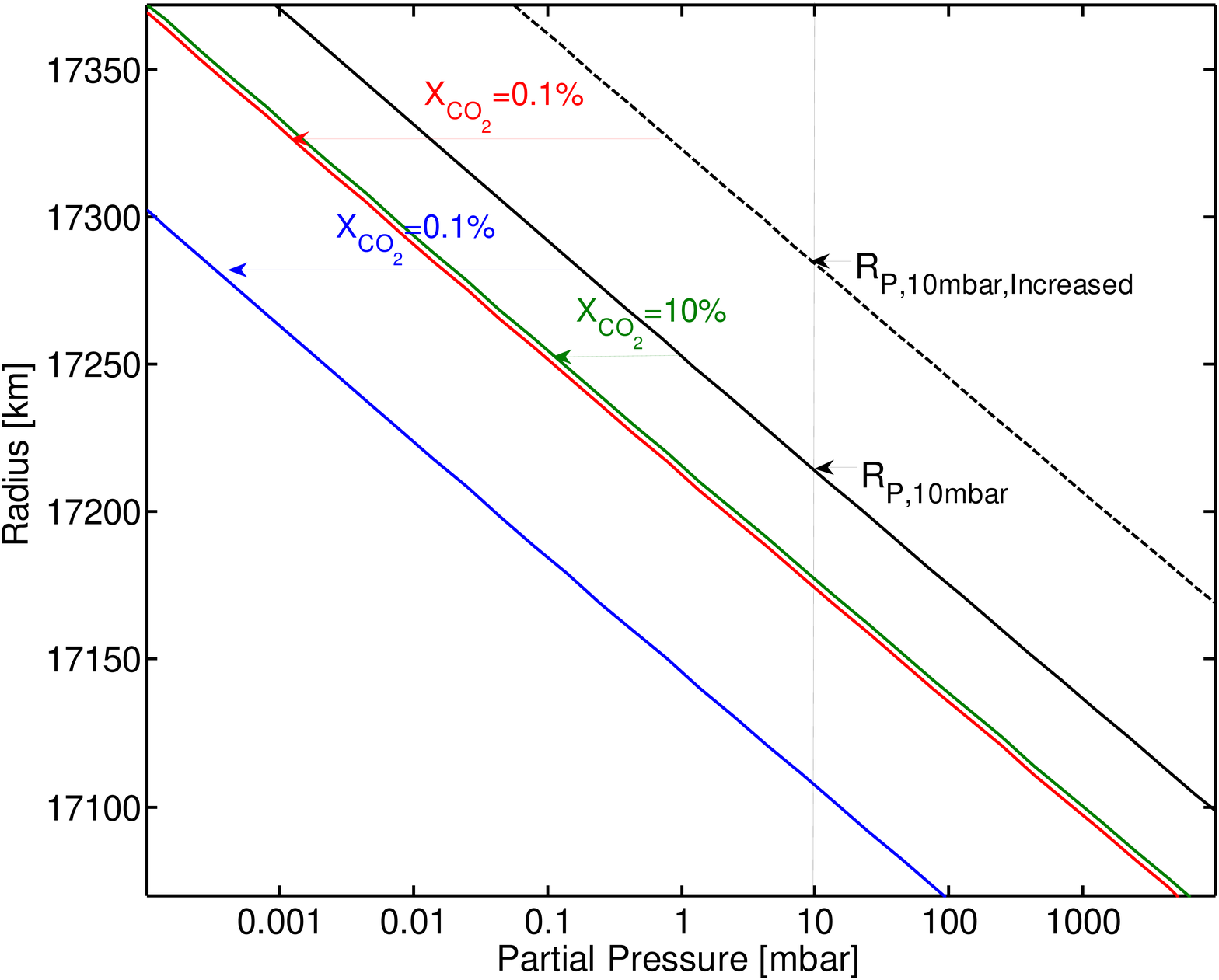}}\hfill{}

\selectlanguage{english}%
\caption{Degeneracy between the absorber mixing ratio, $X_{\mathrm{CO_{2}}}$,
and the planetary radius at the reference pressure level, $R_{P,10}$.
The left panel illustrates the modeled transmission spectra in the
$4.3\mathrm{\mu m}$ $\mathrm{CO_{2}}$ absorption feature for two
different atmospheric compositions. The atmospheric composition of
scenario 1 (red) is 10\% $\mathrm{CO_{2}}$ and 90\% $\mathrm{N_{2}}$.
For the same planetary radius, the transit depth in the absorption
feature of scenario 2 (blue; 0.1\% $\mathrm{CO_{2}}$ and 99.9\% $\mathrm{N_{2}}$)
is lower by $\mathrm{\sim100\, ppm}$ across the entire feature. Increasing
the planetary radius for scenario 2, however, leads to a transmission
spectrum (green) that closely resembles scenario 1. As a result of
this degeneracy between $X_{\mathrm{CO_{2}}}$ and $R_{P,10}$, the
mixing ratio of $\mathrm{CO_{2}}$ cannot be determined to within
several orders of magnitude even for low-noise observations of the
feature. The right panel shows the total pressures for planets with
two different planetary radii (black) and the partial pressure of
$\mbox{C\ensuremath{O_{2}}}$ as a function of the distance from the
planetary center (colors match left panel). Two atmospheres with different
absorber mixing ratios (red and green) can have the same partial pressure/number
density as function of distance from the planetary center leading
to similar absorption features. \label{fig:Degeneracy-between-mixing}}
\selectlanguage{british}
\end{figure*}

\selectlanguage{english}%
The infrared part of the transmission spectrum provides a good tool
to constrain the relative abundances of the molecular absorbers. Constraining
the absolute value of the volume mixing ratios, however, might not
be possible to within orders of magnitude, even with low-noise observations
capturing the shapes of the absorption features because the infrared
part of the spectrum lacks an absolute reference for the transit depth.

The measured transit depths in the absorption features are mainly
related to the number density of the absorbing molecule, $n_{i}\left(r\right)$,
as a function of the radius from center of the planet, $r$. The function
$n_{i}\left(r\right)$, however, provides little useful insight unless
we are able to determine a surface radius or the number density of
a second gas for comparison. In other words, if we do not detect a
surface, then only the mixing ratios of the atmospheric gases have
a meaningful interpretation, not the absolute number densities, because
we are missing an absolute pressure scale. Obtaining the mixing ratios
of an absorbing gas directly by observing the absorption features
of this gas is complicated, however, because different combinations
of the absorber mixing ratio, $X_{i}$, and the planetary radius,
$R_{P,10}$, can lead to the same number density, $n_{i}\left(r\right)$,
and, therefore, to virtually the same absorption feature shape. To
constrain the mixing ratio of a particular gas independently, a reference
for the planetary radius needs to be obtained from a different part
of the spectrum.

\selectlanguage{british}%
\begin{figure*}[tb]
\hfill{}\includegraphics[width=0.5\textwidth]{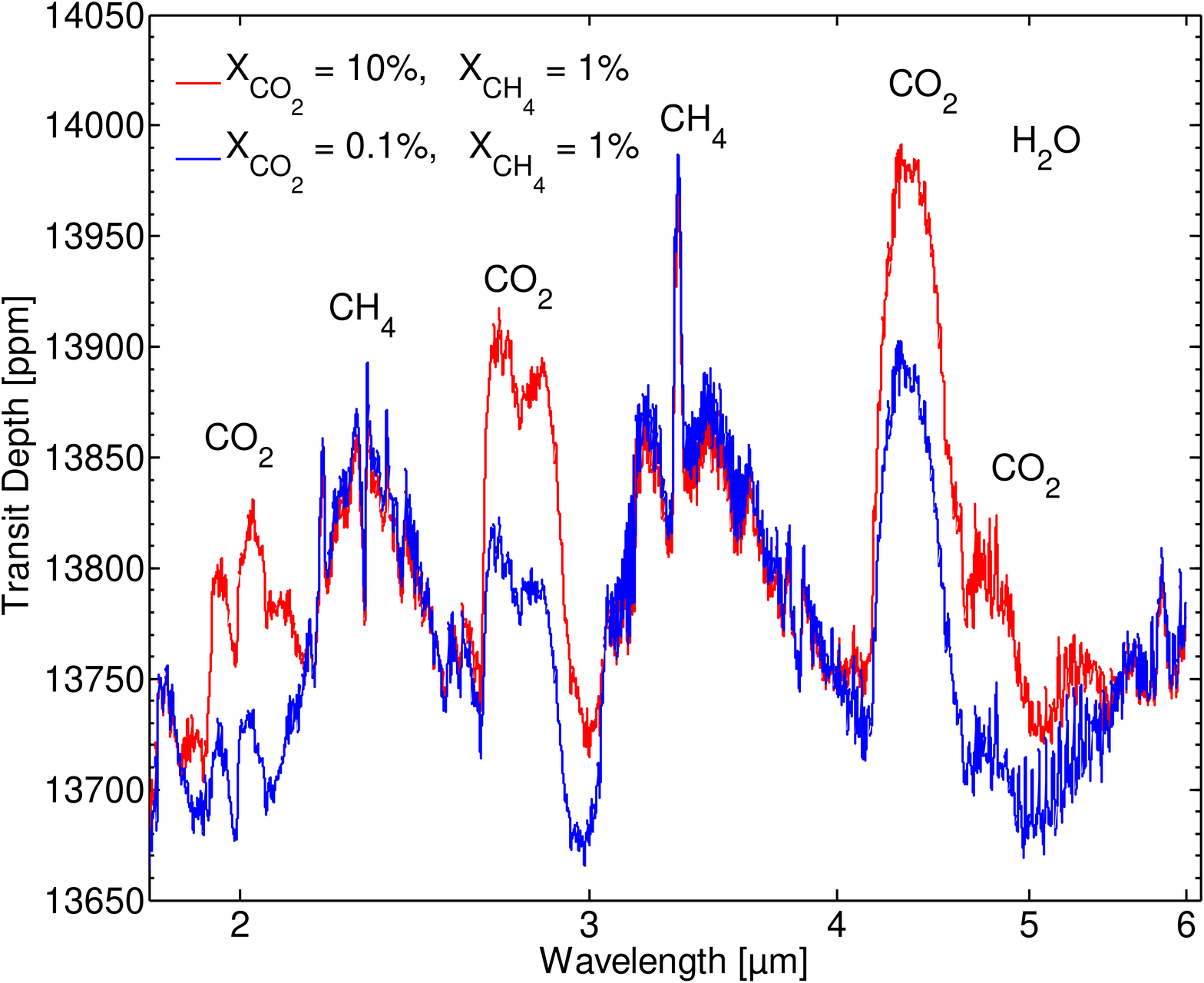}\hfill{}\includegraphics[width=0.5\textwidth]{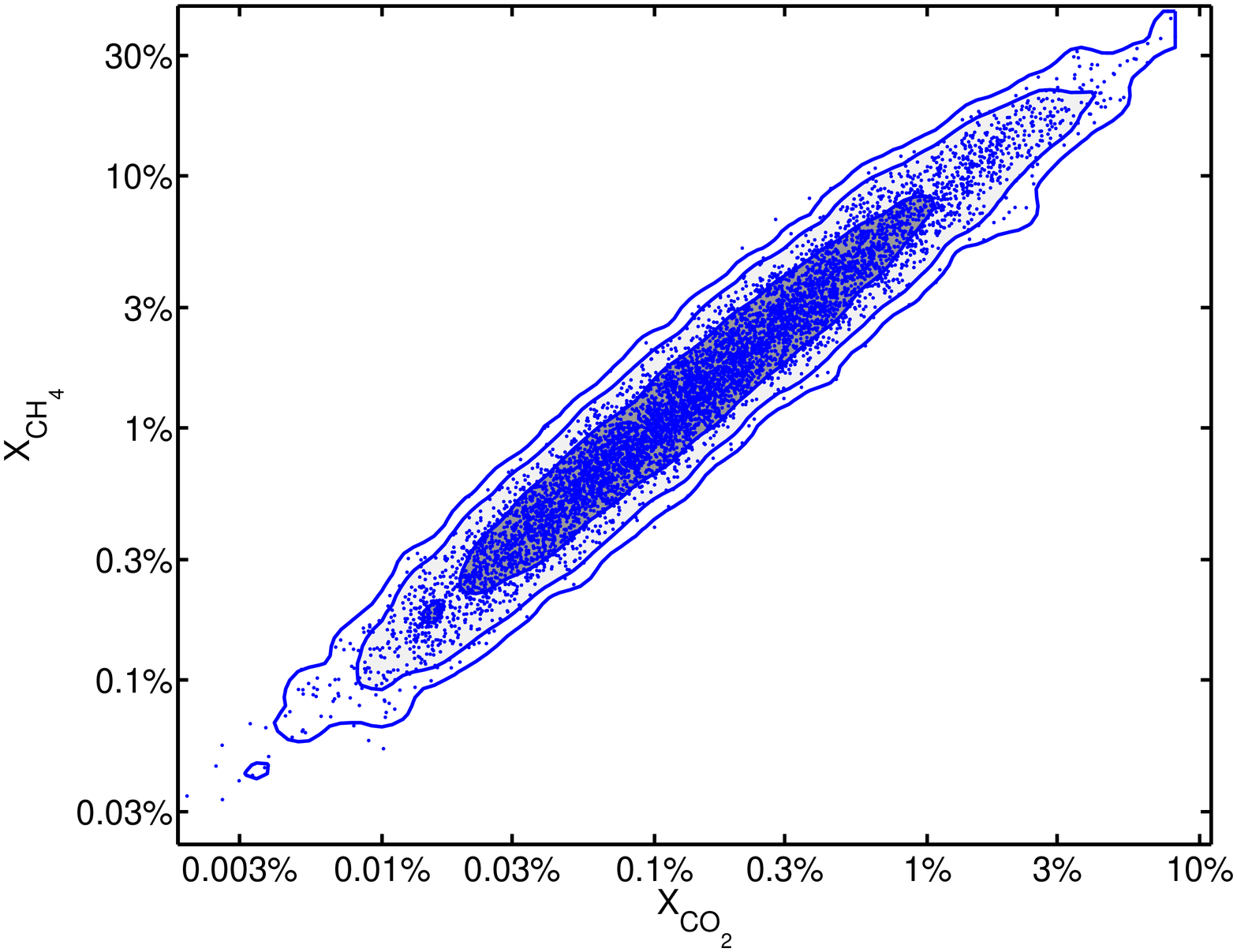}\hfill{}\foreignlanguage{english}{\caption{Constraining the relative abundances of absorbing gases from the near-infrared
spectrum. The left panel shows the transmission spectrum for the atmospheric
scenarios 1 (red, 10\% $\mathrm{CO_{2}}$) and 2 (blue, 0.1\% $\mathrm{CO_{2}}$)
as described in Figure \ref{fig:Degeneracy-between-mixing}, but with
1\% $\mbox{N}_{2}$ replaced by the absorbing gas $\mathrm{CH_{4}}$.
While the $\mathrm{CO_{2}}$ features have a vertical offset between
scenario 1 and scenario 2, the $\mathrm{CH_{4}}$ features are unaffected
by the $\mathrm{CO_{2}}$ mixing ratio. The transit depth in $\mbox{CH}_{4}$
features can, therefore, serve as a relative reference for the $\mathrm{CO_{2}}$
mixing ratio. The right panel shows the two-dimensional marginal posterior
probability distribution of the mixing ratios, $X\mathrm{_{CO_{2}}}$
and $X\mathrm{_{CH_{4}}}$ as retrieved from low-noise synthetic observations
of the near-infrared spectrum ($R$=100, $\sigma_{\left(Rp/R*\right)^{2}}\thickapprox20\,\mbox{ppm},$
$\lambda=2\textrm{-}5\,\mbox{\ensuremath{\mu}m}$) of scenario 1.
The solid lines indicate the 1$\sigma$, 2$\sigma$, and 3$\sigma$
credible regions. Measuring the transit depth in infrared features
of $\mathrm{CO_{2}}$ and $\mathrm{CH_{4}}$ provides good constraints
on the relative abundance ratios of the two gases. The volume mixing
ratios of the gases, however, are strongly correlated. The individual
mixing ratios remain unconstrained across three orders of magnitude
despite robust detections of the infrared features and sufficient
spectral resolution to observe the feature shapes. \label{fig:Correlation-between-abundances:}}
}
\end{figure*}

\selectlanguage{english}%
For a quantitative example, we show the $4.3\,\mathrm{\mu m}$ absorption
feature of $\mathrm{CO_{2}}$ for two different atmospheric compositions
in Figure \ref{fig:Degeneracy-between-mixing}. The compositions are
90\% $\mathrm{N_{2}}$ and 10\% $\mathrm{CO_{2}}$ for scenario 1
and 99.9\% $\mathrm{N_{2}}$ and 0.1\% $\mathrm{CO_{2}}$ for scenario
2. If the planetary radius, $R_{P,10}$, for the two scenarios is
the same, more starlight is blocked in scenario 1 due to the higher
number density $n_{\mathrm{CO_{2}}}\left(r\right)$. The transit depth
inside the spectral features of $\mathrm{CO_{2}}$ is therefore higher
than for scenario 2. If the planetary radius $R_{P,10}$ in scenario
2 is increased by only 70 km ($\approx0.4\%$ ), however, then the
number density $n_{\mathrm{CO_{2}}}\left(r\right)$ in scenarios 2
equals the one in scenario 1, and the absorption feature of scenario
1 closely resembles the absorption feature of the new scenario 2.
The remaining small difference in the transmission spectra is due
to the effect of pressure and temperature on the absorption line broadening.
The effect of changes in line broadening is of secondary order though,
which makes the distinction between scenarios 1 and 2 difficult, even
with extremely low-noise observations. Determining the mixing ratio
of $\mathrm{CO_{2}}$ by observing \textit{only }$\mathrm{CO_{2}}$
features is, therefore, highly impractical. 

A relative reference to break the degeneracy between the planetary
radius and the mixing ratio is provided by the transit depths in absorption
features of different absorbers. Conceptually, this is possible because
a change in the planetary radius affects the absorption features of
both gases equally, while a change in the mixing ratio of one of the
absorbers only affects the features of that absorber. The transit
depth difference between two features of different absorbers is independent
of the planetary radius and only dependent on the relative abundance
ratios of the absorbers and their absorption cross sections in the
features. As the absorption cross section are known from the molecular
databases, comparing the transit depths in features of different absorbers
allows one to constrain the relative abundance of these absorbers.
For a numerical example, we return to our $\mathrm{N}_{2}$-$\mbox{CO}_{2}$
atmosphere and replace 1\% $\mbox{N}_{2}$ by $\mathrm{CH_{4}}$.
In the spectral region around the $3.5\,\mathrm{\mu m}$ $\mathrm{CH_{4}}$
feature (Figure \ref{fig:Correlation-between-abundances:}), the spectrum
remains unaffected by the change in the $\mathrm{CO_{2}}$ mixing
ratio and can serve as a reference to probe the relative abundances
of $\mbox{CO}_{2}$ and $\mbox{CH}_{4}$.

The infrared part of the transmission spectrum covering multiple absorption
features of different molecular species, therefore, provides good
constraints on the relative abundances of the molecular absorbers,
but hardly contains any information on the volume mixing ratios of
the absorbers. Even low-noise NIR observations capturing the shapes
of the absorption features might not constrain the absolute value
of the mixing ratio to within orders of magnitude because the infrared
part of the spectrum provides little information on the abundances
of spectrally inactive gases. 

In our example, the abundance ratio $\frac{X_{\mathrm{CH_{4}}}}{X_{\mathrm{CO_{2}}}}$is
constrained to within a factor of a few at 3$\sigma$, while the volume
mixing ratio of $\mathrm{CH_{4}}$ compatible with the simulated observation
can vary over three orders of magnitude between 0.03\% and 30\%. Note
that the reason for the correlation between $X_{CO_{2}}$and $X_{CH_{4}}$
is \textit{not} the overlap of absorption features of the molecular
species. Overlapping features would cause an \textit{anti}-correlation
between the abundances of the two absorbers.

\subsubsection{Mean Molecular Mass \label{sub:Mean-molecular-mass}}

It has been shown that, for clear atmospheres, measuring the change
in transit depth, $\Delta D$, across spectral features gives an order
of magnitude estimate of the scale height and, therefore, the mean
molecular mass \citep{miller-ricci_atmospheric_2009}. For a general
atmosphere, however, the depth of the absorption features cannot be
used to constrain the mean molecular mass because clouds, hazes, and
a potentially present surface also affect the depths of spectral features.
Here we show for general atmospheres that the value of the mean molecular
mass can be determined by measuring the slope, $\frac{dR_{P,\lambda}}{d\left(ln\sigma_{\lambda}\right)}$
, with which the \textquotedbl{}observed\textquotedbl{} planet radius,
$R_{P,\lambda}$, changes as a function of the extinction cross section,
$\sigma_{\lambda}$, across different wavelengths. In practice, good
observables to independently constrain the mean molecular mass are
(1) the slope Rayleigh scattering signature at short wavelengths,
(2) the relative sizes of strong and weak absorption features of the
same molecule, and (3) the shape of the wings of a strong molecular
absorption feature. 

For the optically thick part of the spectrum, the observed radius
of the planet changes linearly with the logarithm of the extinction
cross section \citep{etangs_rayleigh_2008} and the slope $\frac{dR_{P,\lambda}}{d\text{\ensuremath{\left(ln\sigma_{\lambda}\right)}}}$
is directly related to the atmospheric scale height, 

\begin{equation}
H=\frac{dR_{P,\lambda}}{d\text{\ensuremath{\left(\ln\sigma_{\lambda}\right)}}}.
\end{equation}

A measurement of the observed planet radius, $R_{P,\lambda}$, at
two or more wavelengths with different absorption or scattering cross
sections, $\sigma_{\lambda}$, therefore, permits the determination
of the scale height. Given an estimate of the atmospheric temperature,
e.g. $T\approx T_{eq}$, we can observationally determine an estimate
of the mean molecular mass

\begin{equation}
\mu_{\mathrm{mix}}=\frac{k_{B}T}{g}\left(\frac{dR_{p,\lambda}}{d\text{\ensuremath{\left(\ln\sigma_{\lambda}\right)}}}\right)^{-1}\times\left(1\pm\frac{\delta T}{T}\right),
\end{equation}

where the factor $\left(1\pm\frac{\delta T}{T}\right)$ accounts for
the inherent uncertainty due to the uncertainty, $\delta T$, in modeling
the atmospheric temperature, $T$, at the planetary radius $r=R_{P,\lambda}$
(Appendix). Even if the uncertainty in the temperature estimate is
several tens of percents of the face value, we will find useful constraints
on the mean molecular mass because the mean molecular mass varies
by a factor on the order of $8-20$ between hydrogen-dominated atmospheres
and atmospheres mainly composed of $\mbox{\ensuremath{H_{2}}O}$,
$\mbox{\ensuremath{N_{2}}}$, or $\mbox{C\ensuremath{O_{2}}}$.

The most straightforward way to determine the mean molecular mass
is to measure the slope of the Rayleigh scattering signature at short
wavelengths. The Rayleigh scattering coefficient varies strongly with
wavelength as $\sigma\left(\lambda\right)\varpropto\lambda^{-4}$
. From $\sigma\left(\lambda\right)\varpropto\lambda^{-4}$, we obtain

\begin{equation}
\mu_{mix}=\frac{4k_{B}T}{gR_{*}}\frac{\ln\left(\frac{\lambda_{1}}{\lambda_{2}}\right)}{\left(\frac{R_{p}}{R_{*}}\right)_{\lambda_{2}}-\left(\frac{R_{p}}{R_{*}}\right)_{\lambda_{1}}}\times\left(1\pm\frac{\delta T}{T}\right),\label{eq:H-2}
\end{equation}

\begin{figure}[tb]
\includegraphics[clip,width=1\columnwidth]{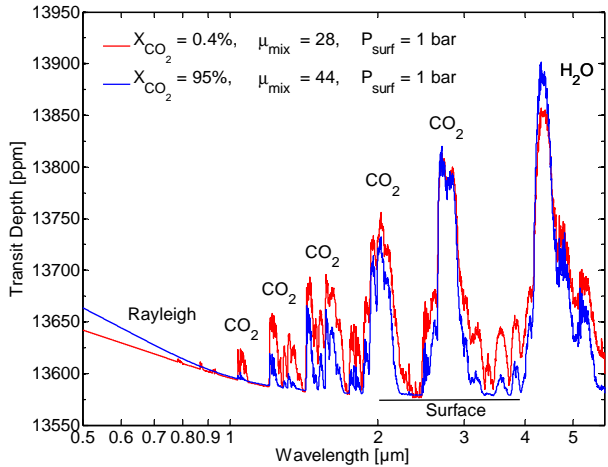}

\caption{Visible to near-infrared transmission spectra for two atmospheric
scenarios with similar absorption feature sizes. The first scenario
(blue) is a $\mathrm{N_{2}}$-rich atmosphere (0.15\% $\mathrm{CO_{2}}$,
99.85\% $\mathrm{N_{2}}$). The second scenario (red) is $\mbox{CO}_{2}$
dominated (95\% $\mathrm{CO_{2}}$ and 5\% $\mathrm{N_{2}}$). Despite
the different mean molecular mass (28 vs. 44) the infrared absorption
feature sizes are similar because the difference in the vertical extent
of the atmosphere due to the different scale heights is compensated
for by the difference in the total amount of the absorbing $\mathrm{CO_{2}}$
gas. A reliable way to determine the scale height is to measure either
the Rayleigh scattering slope $\frac{dR_{P,\lambda}}{d\text{\ensuremath{\left(\ln\lambda\right)}}}$
at short wavelengths, the slope $\frac{dR_{P,\lambda}}{d\text{\ensuremath{\left(ln\sigma_{\lambda}\right)}}}$
in strong absorption features, or the relative feature depths between
strong and weak features of the same molecule. The lower mean molecular
mass atmosphere (blue) shows a steeper Rayleigh scattering slope,
larger differences in the $\mbox{CO}_{2}$ absorption features depth,
and narrower features than the higher mean molecular mass atmosphere
(red). \label{fig:Slope-Effect-of}}
\end{figure}

Measuring the transit depth $D_{\lambda}=\left(\frac{R_{p}}{R_{*}}\right)_{\lambda}^{2}$
at two different wavelengths $\lambda_{1}$ and $\lambda_{2}$ that
are dominated by Rayleigh scattering, therefore, provides the mean
molecular mass. For a quantitative example, we show the transmission
spectra of a $\mathrm{CO_{2}}$-dominated atmosphere (95\% $\mathrm{CO_{2}}$+
5\% $\mathrm{N_{2}})$ and a $\mathrm{N_{2}}$-dominated atmosphere
with small amount of $\mathrm{CO_{2}}$ as the only absorber (0.15\%
$\mathrm{CO_{2}}$, 99.85\% $\mathrm{N_{2}}$) in Figure \ref{fig:Slope-Effect-of}.
Despite the difference in mean molecular mass, the feature depths
are similar due to the different total amounts of the absorber $\mathrm{CO_{2}}$;
thus, the feature depth cannot be used to determine the scale height.
The Rayleigh slope at short wavelength ($\lambda<0.8\mu\mathrm{m}$),
however, is only affected by the scale height and can serve as a good
measure of the mean molecular mass.

A second way of constraining the mean molecular mass is based on analyzing
the detailed shape of the wing and core of spectral features. The
absorption cross section varies strongly from the center to the outer
wings. Measuring the detailed shape of a spectral feature at sufficient
spectral resolution, therefore, probes a large range of cross sections
and allows the constraint of the mean molecular mass. In our example,
the detailed shape of the $4.3\,\mu\mbox{m}$ $\mathrm{CO_{2}}$ feature
shows the difference between the scenarios (Figure \ref{fig:Slope-Effect-of}).
For smaller mean molecular mass, the feature is higher at the center
with narrow wings, while the large mean molecular mass leads to broader
features. The measurement of this difference requires at least a moderate
spectral resolution ($R\sim50$) and a high signal-to-noise ratio
(S/N).

A third way to probe the mean molecular mass is to quantitatively
compare the broadband transit depths in different spectral features
of the same absorber. Again, we probe the planetary radius at wavelengths
for which the cross sections are different: strong absorption features
have large absorption cross sections, while weaker features of the
same absorber have smaller cross sections. A quantitative comparison
of the depths of individual features therefore provides the gradient
$\frac{dR_{P,\lambda}}{d\text{\ensuremath{\left(ln\sigma_{\lambda}\right)}}}$
and constrains the scale height and mean molecular mass. For atmospheres
with small mean molecular masses, the gradient $\frac{dR_{P,\lambda}}{d\text{\ensuremath{\left(ln\sigma_{\lambda}\right)}}}$
is large, resulting in greater differences in the transit depths between
the strong and the weak features (Figure \ref{fig:Slope-Effect-of}).

\subsubsection{Volume Mixing Ratios of the Atmospheric Constituents \label{sub:Refractive-index-of}}

The primary quantities affected by the abundance of spectrally inactive
gases are the estimate of the mean molecular mass, $\mu_{\mathrm{mix}}$,
discussed in Section \ref{sub:Mean-molecular-mass} and the transit
depth offset of the molecular Rayleigh scattering slope, $D_{\mathrm{Rayl}}$.
Combining the information on $\mu_{\mathrm{mix}}$ and $D_{\mathrm{Rayl}}$
with the constraints on the relative abundances of the absorbers from
the NIR spectrum (Section \ref{sub:Abundance-ratios-of}) provides
unique constraints on the volume mixing ratios of both the spectrally
inactive gases and molecular absorbers.

The atmospheres of Jupiter-sized planets present a simplified case
for atmospheric retrieval. From their radius and mass measurements,
we can conclude that they have accreted a hydrogen-dominated atmosphere,
thus we know the mean molecular mass a priori. Constraining the volume
mixing ratios of the molecular species in the atmosphere, nonetheless,
requires the observation of the transit depth offset of the molecular
Rayleigh scattering slope, $D_{\mathrm{Rayl}}$ at short wavelength.

Neglecting for now the effect of refractive index variations between
different gas mixtures, the transit depth offset of the molecular
Rayleigh scattering slope, $D_{\mathrm{Rayl}}$, is only a function
of planetary radius at the reference pressure, $R_{P,10}$. The transit
depth offset of the Rayleigh scattering signature can, therefore,
serve as a reference transit depth to obtain an absolute scale for
the atmospheric pressure and to determinethe volume mixing ratios
of the absorbers from the absorption features in the NIR. In general,
atmospheres rich in absorbing gases will show transmission spectra
for which the transit depth in the Rayleigh scattering signature is
small with respect to the transit depth in the NIR, while atmospheres
dominated by spectrally inactive gases will show transmission spectra
that have a strong Rayleigh scattering signatures and absorption features
in the NIR at a lower transit depth levels.

Obtaining the absolute abundances for all relevant absorbing gases
enables us to constrain the total mixing ratio of the spectrally inactive
gases to be $X_{\mathrm{inactive}}=1-\sum_{i=1}^{n}X_{i}$, where
$n$ is the number of absorbers in the atmosphere. Conceptually, the
estimate of the mean molecular mass, $\mu_{\mathrm{mix}}$, can then
be used to determine the individual mixing ratios of the spectrally
inactive components, $\mbox{N}_{2}$ and $\mbox{H}_{2}+\mbox{He}$.
We obtain the volume mixing ratios of $\mathrm{N}_{2}$ and primordial
gas from

\begin{equation}
\mu_{mix}=\mu_{N_{2}}X_{N_{2}}+\mu_{H_{2}+He}\left(X_{inactive}-X_{N_{2}}\right)+\sum_{i=1}^{n}\mu_{i}X_{i}
\end{equation}

and

\begin{equation}
X_{\mathrm{H_{2}+He}}=X_{\mathrm{inactive}}-X_{\mathrm{N_{2}}}.
\end{equation}

Individual constraints on $\mbox{H}_{2}$ and He are not possible
because only two spectrally inactive gases can be fit. Three or more
individual spectrally inactive components inherently lead to degeneracy
because the same mean molecular mass of the spectrally inactive gases
can be obtained by different combinations of the mixing ratios of
the gases.

In reality, the effective refractive index of the gas mixture varies
depending on the composition and affects the transit depth offset
of the Rayleigh scattering signature (Section \ref{sub:Opacities}).
When simultaneously retrieving the mixing ratios of all gases, however,
we also determine the refractive index in the process because the
refractive index is a direct function of only the mixing ratios and
not an additional unknown.

\subsubsection{Surface Pressure \label{sub:Surface-Pressure}}

We can discriminate between a thick, cloud-free atmosphere and an
atmosphere with a surface, where the surface is either the ground
or an opaque cloud deck. For atmospheres with an upper surface at
pressures lower than $P_{\mathrm{surf}}\lesssim\mathrm{100\, mbar\ldots5\, bar}$
(depending on composition), we can quantitatively constrain the pressure
at this surface. For a thick atmosphere, we can identify a lower limit
on the surface pressure.

A surface strongly affects the part of the spectrum without absorption
features while having only a weak or negligible effect on the part
of the transmission spectrum with strong molecular absorption or scattering.
In the spectral regions with weak absorption, a thin atmosphere has
a relatively constant continuum because the surface cuts off the grazing
light beams at a radius that is independent of the wavelength \citep{des_marais_remote_2002,ehrenreich_transmission_2006}.
A thick atmosphere without a surface lacks a flat continuum. 

Conceptually, the optically-thick regions of the spectrum, those for
which the transit depth is independent of the surface pressure, constrain
the mixing ratios of the molecular species in the atmosphere as described
in Section \ref{sub:Atmospheric-Composition}. The surface pressure
can then be determined from the transit depths in the parts of the
spectrum in which absorption and scattering are weak (Figure \ref{fig:Flow-Chart}).
For a noise-free spectrum, the strongest constraint on the surface
pressure is provided by the minimum transit depth, $D_{\mathrm{min}}$,
measured across the spectrum. The minimum transit depth determines
the deepest pressure level for which light is transmitted through
the atmosphere and, therefore, provides a lower limit on the surface
pressure. In practice, the retrieval of the mixing ratios, surface
pressure, and other parameters is performed simultaneously based on
the information in the entire spectrum.

\begin{figure}[tb]
\includegraphics[clip,width=1\columnwidth]{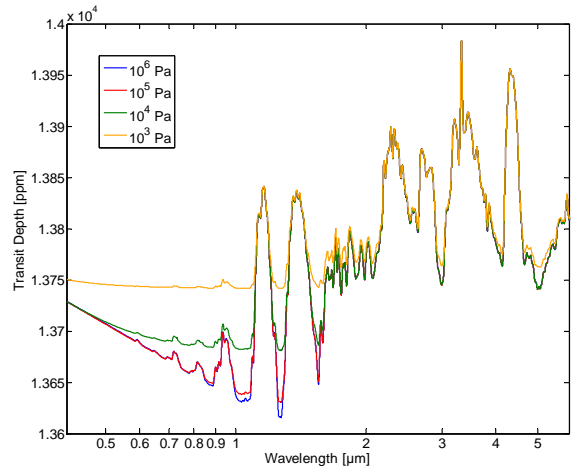}\caption{Effect of the surface pressure on the transmission spectra of exoplanets.
The transmission spectra of model atmospheres with 99\% $\mathrm{N_{2}}$
and 1\%$\mathrm{CO_{2}}$ are depicted for four different surface
pressures. The radius, $R_{P,10}$, at which the atmospheric pressure
is 10 mbar is set to the same value for all four. With $R_{P,10}$
set to the same value, the transit depths in the strong $\mathrm{CO_{2}}$
absorption bands (e.g., around 2.7, 3.3, and 4.3$\mathrm{\mu m}$)
are independent of the surface pressure since the grazing star light
at these wavelengths does not penetrate to lower layers of the atmosphere.
Conversely, the parts of the spectrum ($<1.6\,\mu\mathrm{m}$) with
little molecular absorption and scattering show a strong dependence
on the surface pressure. Combining information from parts of the spectrum
that are sensitive to surface pressure and parts of the spectrum that
are insensitive to surface pressure enables one to find independent
constraints on atmospheric composition and surface pressure. At sufficiently
high surface pressures, even the spectral regions with low absorption
cross sections become optically thick for a grazing light beam and
the complete spectrum becomes insensitive to further increases in
the surface pressure. For these thick atmospheres, only a lower limit
on the surface pressure can be found. \label{fig:PsurfEffect}}
\end{figure}

Taking the example of a $\mathrm{N_{2}}$-$\mathrm{CO_{2}}$ atmosphere
(Figure \ref{fig:PsurfEffect}), the shape of the spectral features
in the 2-6 $\mathrm{\mu m}$ range is mostly unaffected by changes
in surface pressure, as long as the surface pressure is higher than
$100\,\mathrm{mbar}$. For exquisite data, the composition of the
atmosphere can, therefore, be retrieved from the 2 to $\mathrm{6\,\mu m}$
range independently of the surface pressure. Conversely, the spectral
region between 0.5 and 2 $\mathrm{\mu m}$ is strongly affected by
surface pressure, but the effects of surface pressure and mixing ratios
are usually degenerate. Taking the retrieved mixing ratios from the
part of the spectrum unaffected by surface pressure, allows a unique
determination of the surface pressure.

\subsection{Numerical Results\label{sub:Quantitative-constraints-on}}

\subsubsection{Constraints on Composition}

In this section, we present numerical results for synthetic \textit{JWST}
NIRSpec observations of the transmission spectrum of the super-Earth
GJ~1214b. In all three atmospheric scenarios studied, we find that
the analysis of moderate spectral resolution ($R\thickapprox100)$
transmission spectra covering the spectral range between $0.6$ and
$5\,\mu\mbox{m}$ can provide narrow probability posterior distributions
for all absorbing gases with mixing ratios of several ppm or higher.
(Figures \ref{fig:OutputPlot}-\ref{fig:H2_Marginalized-posterior-distribut}).
The well-constrained probability distributions allow a direct inference
of the most likely estimate and credible regions (Bayesian equivalent
to confidence intervals) for the mixing ratios of the individual molecular
species. Spectrally inactive gases can also be constrained if their
abundances are sufficient to affect the mean molecular mass and the
Rayleigh scattering signature at short wavelengths.

For a given transmission spectrum, the relative uncertainties in the
mixing ratios $\frac{\Delta X_{i}}{X_{i}}$ of absorbing gases (e.g.,
$\mathrm{H_{2}O},$ $\mathrm{CO_{2}}$, $\mathrm{CH_{4}}$) are only
weakly dependent on the absolute values of the mixing ratios. In other
words, minor gases with mixing ratios as low as tens of ppm can be
constrained as well as the major atmospheric constituents (e.g., Figure
\ref{fig:N2Marginalized-posterior-distribut}(a)). The reason for
this is that the long geometric path length of the grazing stellar
light through the atmosphere of the extrasolar planet leads to significant
spectral features in the transmission spectrum even for low-abundance
gases \citep{brown_transmission_2001}.  Increasing the mixing ratio
increases the transit depth across the feature, but the uncertainty
in the observed transit depth and, therefore, the uncertainty on the
logarithm of the mixing ratio remains mostly unchanged. A detection
limit does exist at low abundances, however because overlapping features
of other absorbers may mask the features of extremely low-abundance
gases. If all spectral regions in which the absorber is active are
occupied by stronger features of other absorbers, then only an upper
limit on the mixing ratio of the gas can be found (Figure \ref{fig:H2_Marginalized-posterior-distribut}).

In constrast to the absorbing gases, the uncertainties in the mixing
ratios of spectrally inactive gases (e.g., $\mathrm{N_{2}}$, $\mathrm{H_{2}}$)
are strongly dependent on their mixing ratios. Spectrally inactive
gases affect the transmission spectrum only through changing the mean
molecular mass and changing the transit depth difference between the
Rayleigh scattering signature and the NIR spectrum (Section \ref{sub:Refractive-index-of}).
If the mixing ratio of a spectrally inactive gas is a few tens of
percent or more, the effect of the spectrally inactive gas on the
atmospheric mean molecular mass is strong and, therefore, it is relatively
easy to identify the spectrally inactive gas and constrain its mixing
ratio (Figures \ref{fig:N2Marginalized-posterior-distribut} and \ref{fig:H2_Marginalized-posterior-distribut}).
 For lower mixing ratios, however, only weak constraints or an upper
limit can be placed on the mixing ratios of spectrally inactive gases
because their effect on the spectrum becomes negligible (e.g., Figure
\ref{fig:OutputPlot}). This is particularly true for $\mathrm{N_{2}}$
whose molecular mass (28~u) differs only by a factor of $\sim1.6$
or less from the molecular masses of the most common spectrally active
gases, e.g., $\mathrm{H_{2}O}$ (18~u), $\mathrm{CH_{4}}$ (18~u),
and $\mathrm{CO_{2}}$ (44~u). Constraining the mixing ratio of $\mathrm{H_{2}}$
is achieved down to lower mixing ratios because its molecular mass
is lower than that of most absorbing gases by a factor of six or more.

\subsubsection{Constraints on Surface Pressure}

\begin{figure*}[tb]
\selectlanguage{british}%
\hfill{}\includegraphics[width=0.5\textwidth]{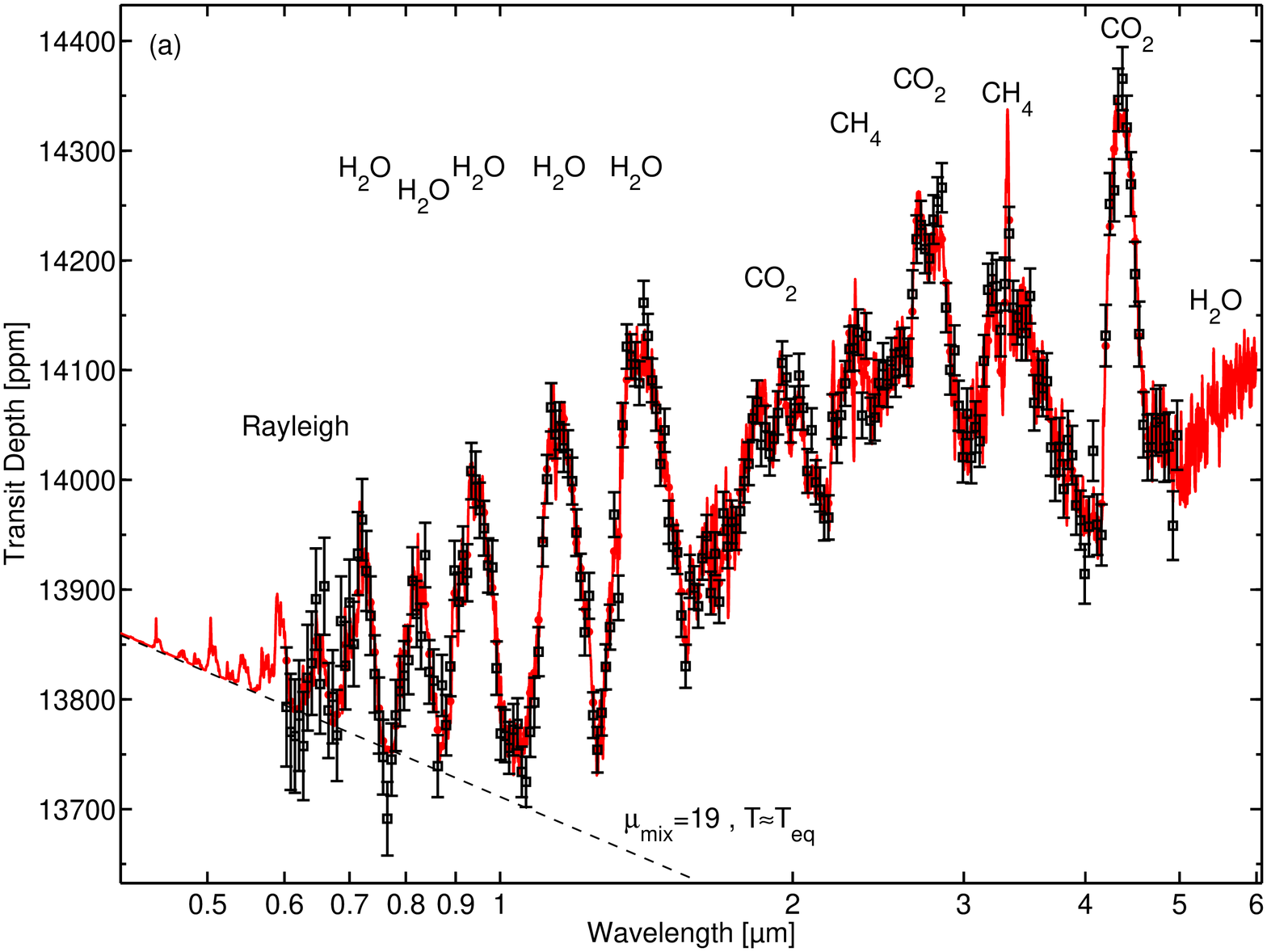}\hfill{}\includegraphics[width=0.5\textwidth]{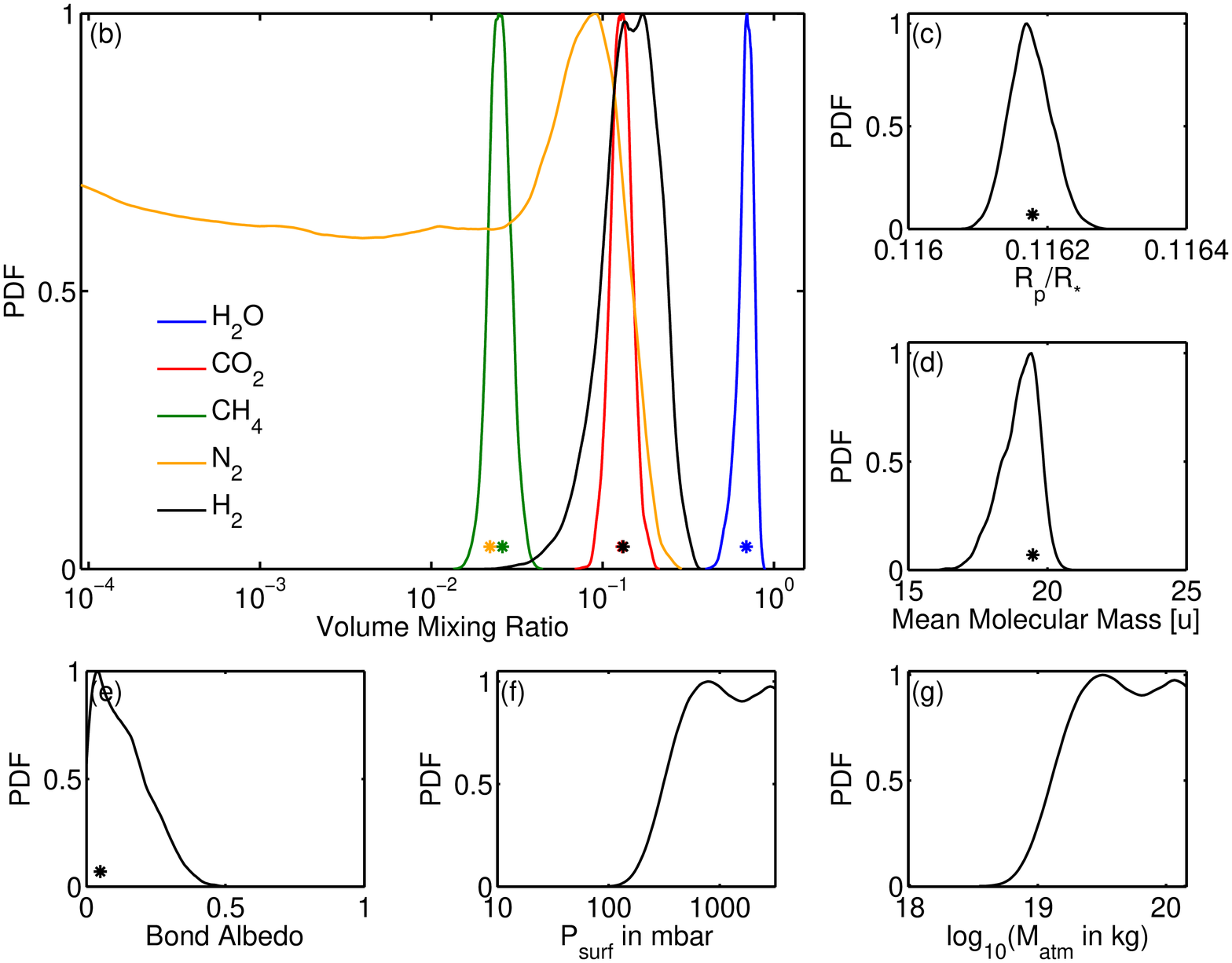}\hfill{}

\selectlanguage{english}%
\caption{Synthetic transit observations and atmospheric retrieval results for
the \textquotedbl{}Hot Halley world\textquotedbl{} scenario for the
super-Earth GJ~1214b. The synthetic observations shown in panel (a)
were simulated considering 10 transit observations with \textit{JWST}
NIRSpec and assuming that observational uncertainties within 20\%
of the shot noise limit are achieved (see Section \ref{sub:Observation-Scenarios}).
The dashed line shows the analytical Rayleigh scattering slope for
comparison. Panels (b)-(g) illustrate the marginalized posterior probability
distribution for the atmospheric parameters retrieved from the synthetic
observations. For illustrative purposes, the distributions are normalized
to a maximum value of 1. The asterisks indicate the values of the
atmospheric parameters used to simulate the input spectrum. The narrow,
single-peaked, posterior probabilities for the mixing ratios of $\mathrm{H_{2}O,}$
$\mathrm{CO_{2}}$, $\mathrm{CH_{4}}$, and $\mbox{H}_{2}$ in panel
(b) indicate that unique constraints on the abundance of these gases
can be retrieved in agreement with the atmospheric parameters used
to simulate the input spectrum. $\mathrm{H_{2}O}$ can be identified
as the main constituent. Only an upper limit on the mixing ratio of
$\mbox{N}_{2}$ can be found because small amounts of the spectrally
inactive $\mbox{N}_{2}$ have a negligible effect on the observed
transmission spectrum. Constraints are also obtained for the surface/cloud-top
pressure and total atmospheric mass above the surface/cloud-top (Panels
(f) and (g)). In this scenario, the atmosphere is cloud-free down
to high pressure levels, thus only a lower bound on the surface pressure
can be found. No upper bound can be inferred as indicated by the posterior
probability distributions approaching the flat prior distribution
at high surface pressures. \label{fig:OutputPlot}}
\end{figure*}

\begin{figure*}[tb]
\selectlanguage{british}%
\hfill{}\includegraphics[width=0.5\textwidth]{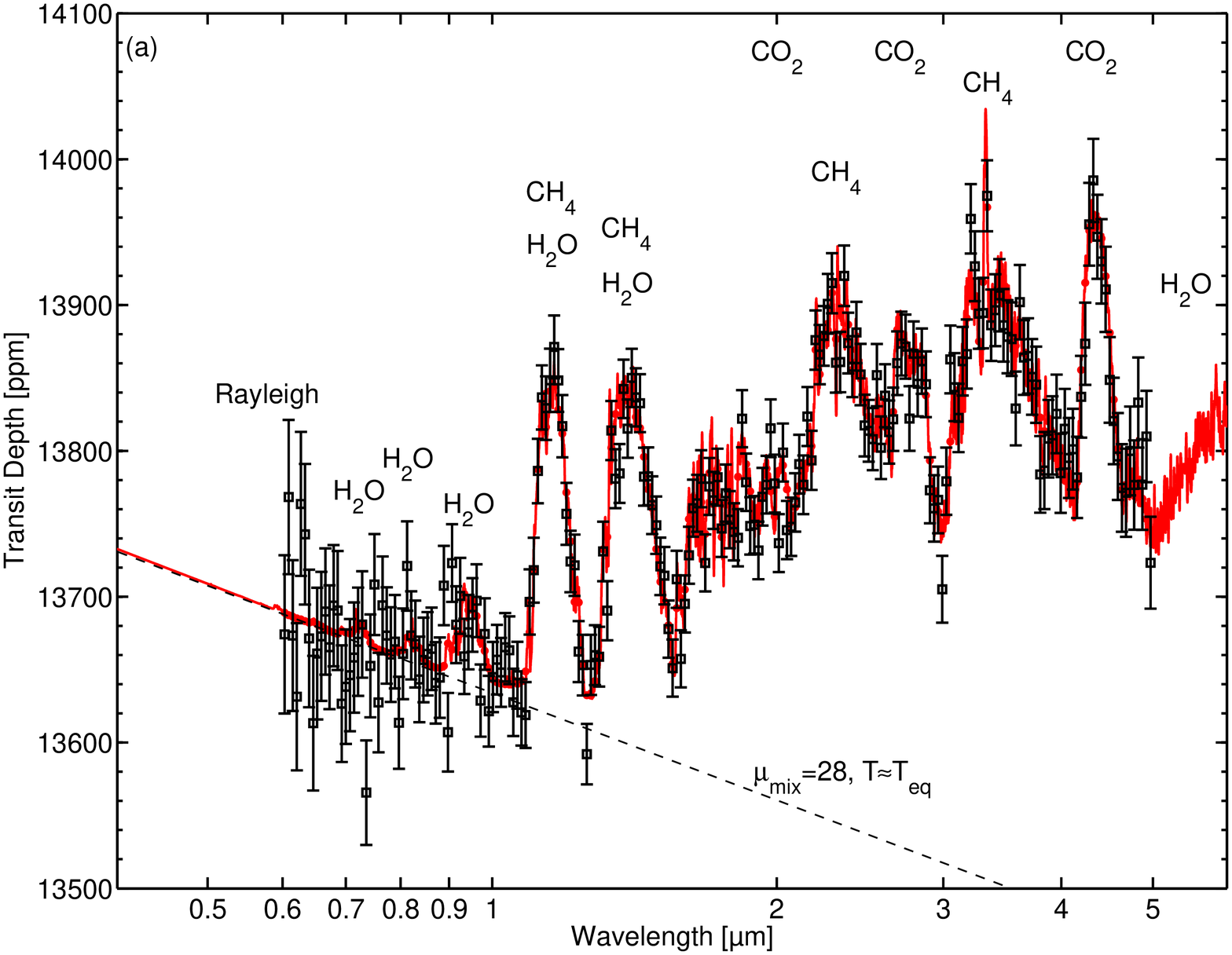}\hfill{}\includegraphics[width=0.5\textwidth]{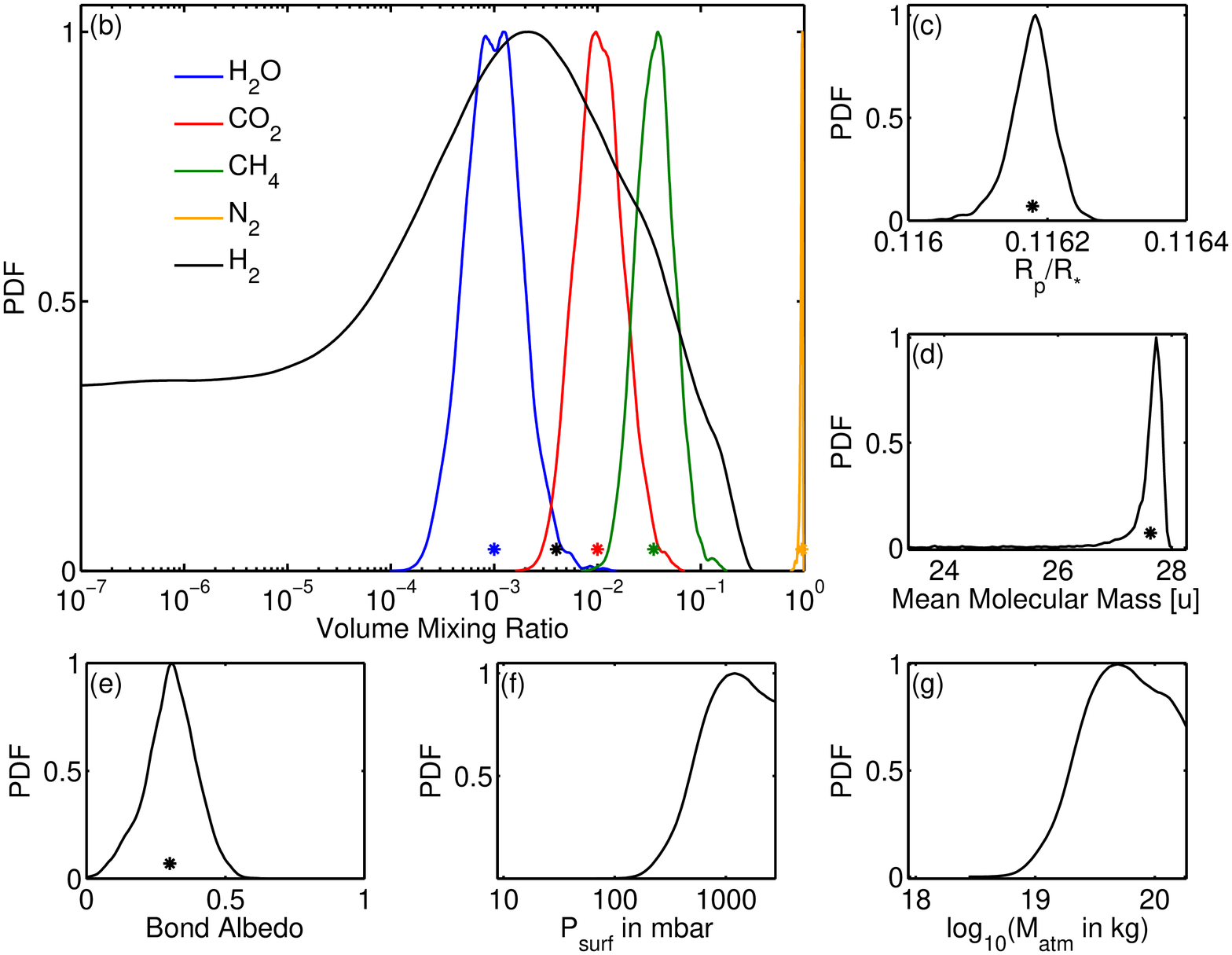}\hfill{}

\selectlanguage{english}%
\caption{Synthetic transit observations and atmospheric retrieval results for
the \textquotedbl{}Hot nitrogen-rich world\textquotedbl{} scenario
of a super-Earth with the physical properties of GJ~1214b. The panels
identities are identical to Figure \ref{fig:OutputPlot}. Observational
errors were modeled for 10 transit observations with \textit{JWST}.
The narrow posterior distributions for the mixing ratios of $\mathrm{H_{2}O,}$
$\mathrm{CO_{2}}$, $\mathrm{CH_{4}}$, and $\mbox{N}_{2}$ indicate
that unique constraints on the abundance of these gases can be retrieved
in agreement with the atmospheric parameters used to simulate the
input spectrum. $\mathrm{N_{2}}$ can be identified to be the main
constituent of the atmosphere due to its effect on the mean molecular
mass and the Rayleigh signature. While atmospheric models with $X_{\mbox{H}_{2}}\approx0.1\ldots1\%$
are favored by the synthetic observations, atmospheric models with
$X_{\mbox{H}_{2}}\rightarrow0$ retain a significant probability and
no lower bound on $X_{\mbox{H}_{2}}$ can be found. The most likely
value for the surface pressure is in agreement with the surface pressure
parameter used to simulate the input spectrum, suggesting that the
atmosphere is optically thin at some wavelengths. The synthetic observations
are not sufficient, however, to find a statistically significant upper
limit on the surface pressure and fully exclude a thick envelope.
\label{fig:N2Marginalized-posterior-distribut}}
\end{figure*}

\begin{figure*}[tb]
\selectlanguage{british}%
\hfill{}\includegraphics[width=0.5\textwidth]{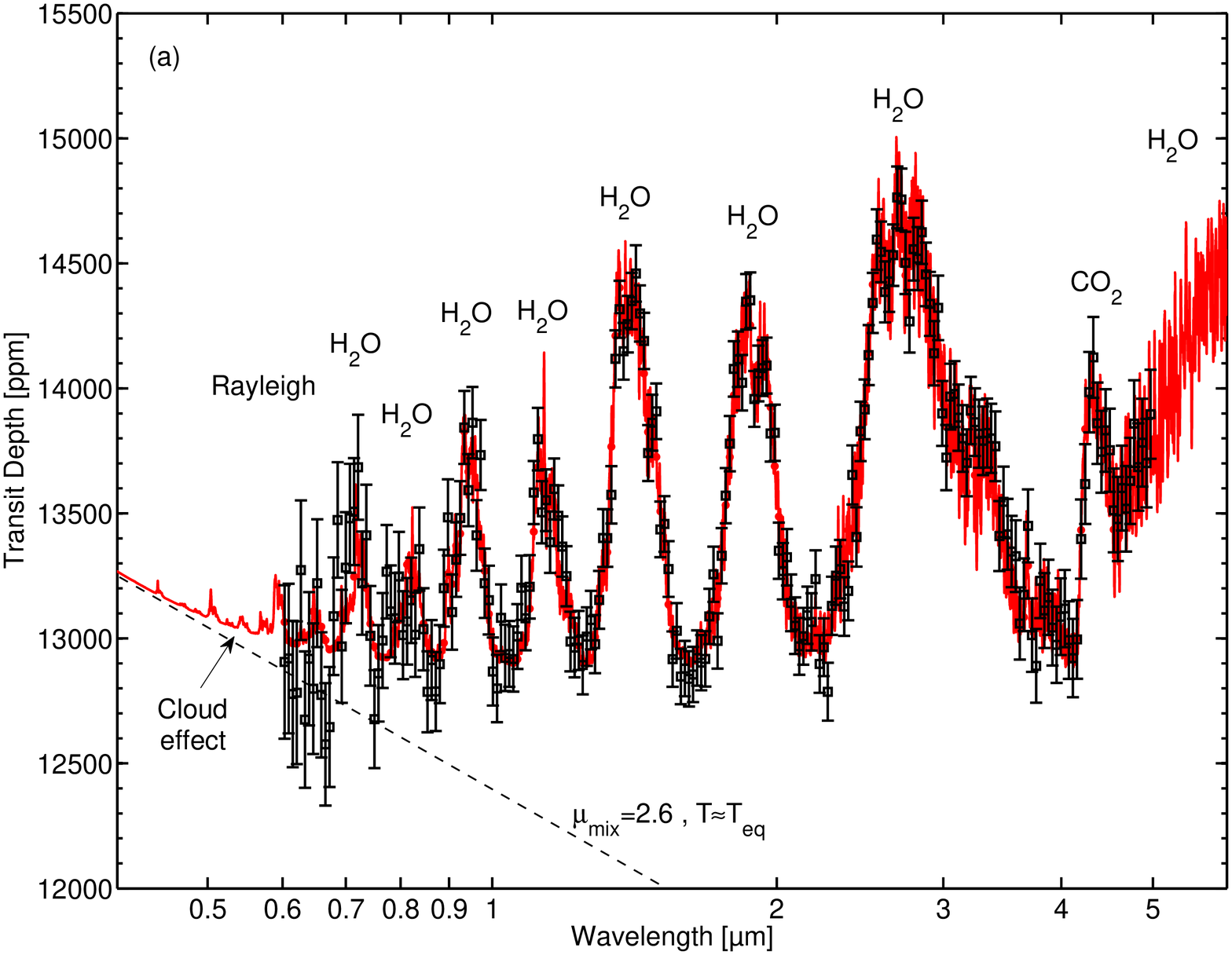}\hfill{}\includegraphics[width=0.5\textwidth]{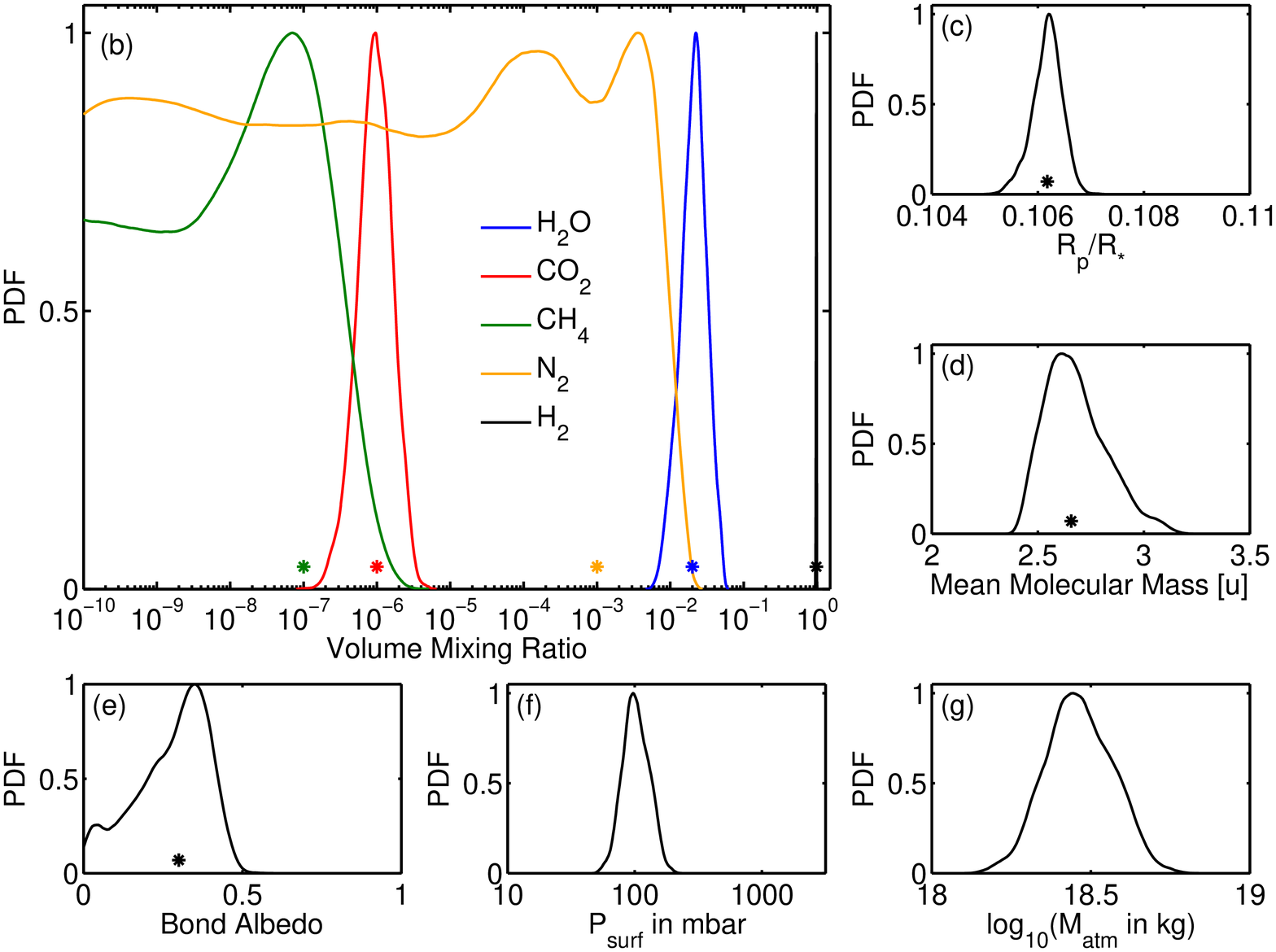}\hfill{}

\selectlanguage{english}%
\caption{Synthetic transit observations and atmospheric retrieval results for
the \textquotedbl{}Hot mini-neptune\textquotedbl{} scenario of a super-Earth
with the physical properties of GJ~1214b. The panels identities are
identical to Figure \ref{fig:OutputPlot}. Observational errors were
modeled for a single transit observation with \textit{JWST} NIRSpec.
Note the difference in the scale of the transit depth axis compared
to Figures \ref{fig:OutputPlot} and \ref{fig:N2Marginalized-posterior-distribut}.
The narrow posterior distributions for the mixing ratios of $\mathrm{H_{2}O,}$
$\mathrm{CO_{2}}$, and $\mathrm{H_{2}}$ indicate that unique constraints
on the abundance of these gases can be retrieved in agreement with
the atmospheric parameters used to simulate the input spectrum. Based
on the low mean molecular mass, $\mathrm{H_{2}}$ can clearly be identified
as the main constituent of the atmosphere. $\mathrm{N_{2}}$ mixing
ratios larger than a few percent can be excluded. An upper limits
at the ppm level can be found for the mixing ratio of $\mathrm{CH_{4}}$.
A surface (here: due to the opaque cloud deck) can be identified at
a pressure level between 65 and 150 mbar with 3$\sigma$ confidence.
\label{fig:H2_Marginalized-posterior-distribut}}
\end{figure*}

In the retrieval output, a thin atmosphere with a surface and a thick,
cloud-free atmosphere show distinct posterior probability distributions
for the surface pressure parameters. For atmospheres that are thin
or have an upper cloud deck at low pressure levels, the probability
distribution resembles a well-constrained, single-modal distribution
(Figure \ref{fig:H2_Marginalized-posterior-distribut}(d)). For thick
atmospheres that lack an observable surface, only an upper limit to
the surface pressure can be retrieved (Figure \ref{fig:OutputPlot}(d))The
posterior probability of the surface pressure plateaus toward high
pressures, indicating that further increases in surface pressure lead
to equally likely scenarios. We emphasize that, for a terrestrial
planet, the two scenarios of a thin atmosphere with a solid surface
or a thick atmosphere with an opaque cloud deck are not distinguishable
from the transmission spectrum.

\subsubsection{Effect of Unobserved Temperature}

\begin{figure*}[tb]
\selectlanguage{british}%
\noindent \begin{centering}
\includegraphics[clip,width=0.8\textwidth]{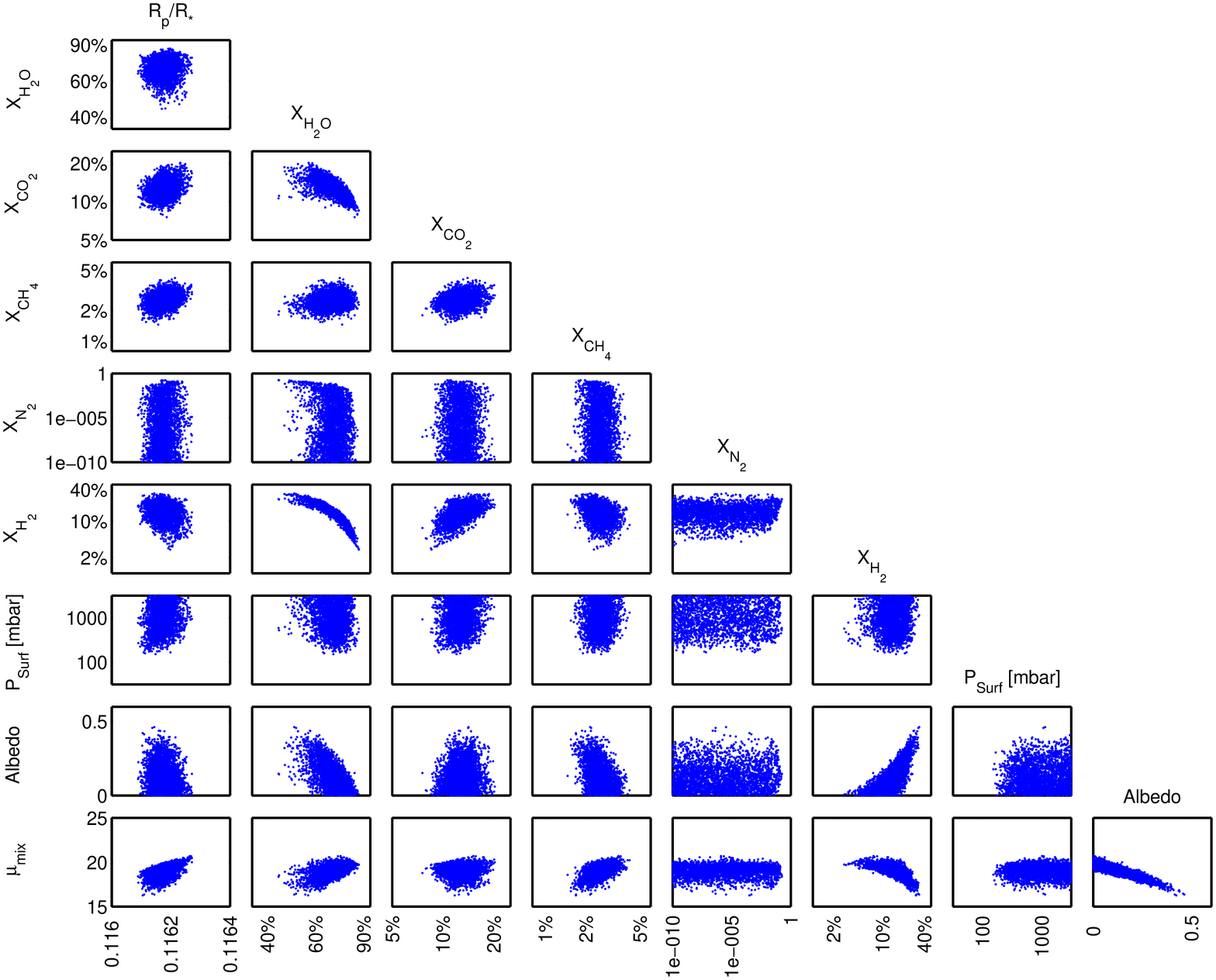}
\par\end{centering}

\selectlanguage{english}%
\noindent \centering{}\caption{Two-dimensional marginalized probabilities for pairs of atmospheric
properties for simulated \textit{JWST} NIRSpec observations of the
\textquotedbl{}Hot Halley world\textquotedbl{} scenario for GJ~1214b.
The synthetic observation used for the atmospheric retrieval is illustrated
in Figure \ref{fig:OutputPlot}. For observations that cover all $n+4$
observables discussed in Section \ref{sub:Atmospheric-Composition},
the posterior distribution of the atmospheric parameters retrieval
lacks degeneracies or strong correlations that would keep individual
parameters unconstrained over orders of magnitude (Figure \ref{fig:2-D-marginalized-probability}).
Planetary albedo and the mean molecular mass show a correlation because
different combinations of atmospheric temperature and mean molecular
mass may lead to the same scale height and, therefore, to similar
spectral feature shapes. \label{fig:2-D-marginalized-probability}}
\end{figure*}

An inherent correlation arises between the planetary albedo and the
mean molecular mass (Figure \ref{fig:2-D-marginalized-probability})
if no direct measurements of the planetary temperature or the planetary
albedo are available. While the correlation does not lead to uncertainties
of individual parameters that range over orders of magnitude, it can
be the dominant source of uncertainty on the composition if small
error bars are achieved for the observations of the primary transit,
but no direct measurements of the brightness temperature or the planetary
albedo are available from secondary eclipse observations .

The reason for the correlations between mean molecular mass and albedo
is that the primary observables for the mean molecular mass (see Section
\ref{sub:Mean-molecular-mass}) constrain the scale height rather
than the mean molecular mass directly. Given the observational constraints
for the scale height, different combinations of the atmospheric temperature
and mean molecule mass may agree equally well with the scale height
constraints imposed by the spectrum. The atmospheric temperature,
in turn, is primarily determined by the planetary albedo, giving rise
to the correlation between planetary albedo and mean molecular mass.

A higher mixing ratio of $\mbox{H}_{2}$ lowers the mean molecular
mass without creating new absorption features. An atmosphere with
more $\mbox{H}_{2}$ and less of the main constituent (here: $\mbox{H}_{2}\mbox{O}$)
in conjunction with an increased planetary albedo shows virtually
the same transmission spectrum as the one shown in Figure \ref{fig:OutputPlot}.
As a result, the posterior distribution shows a significant correlation
between $X_{\mathrm{H_{2}}}$ and the Bond albedo as well as between
$X_{\mathrm{H_{2}}}$ and $X_{\mathrm{H_{2}O}}$ (Figure \ref{fig:2-D-marginalized-probability}). 

\subsection{Elemental Abundances}

The ability to constrain the mixing ratios of both the absorbing and
the spectrally inactive gases in the atmosphere provides us with the
opportunity to probe the relative abundances of the volatile elements
H, C, O, and N of the atmospheres of exoplanets. Conceptually, the
retrieval of the elemental abundances in the atmosphere is directly
linked to the retrieval of the molecular mixing ratios, since the
constraints on elemental abundances are derived from the probability
density distribution of the molecular mixing ratios (Section \ref{sub:Output}).
Following the result in the previous subsections, low-noise observations
of moderate to high spectral resolution lead to well-constrained molecular
mixing ratios and therefore also allow determination of well-constrained
elemental abundances.

\begin{figure}[tb]
\includegraphics[clip,width=1\columnwidth]{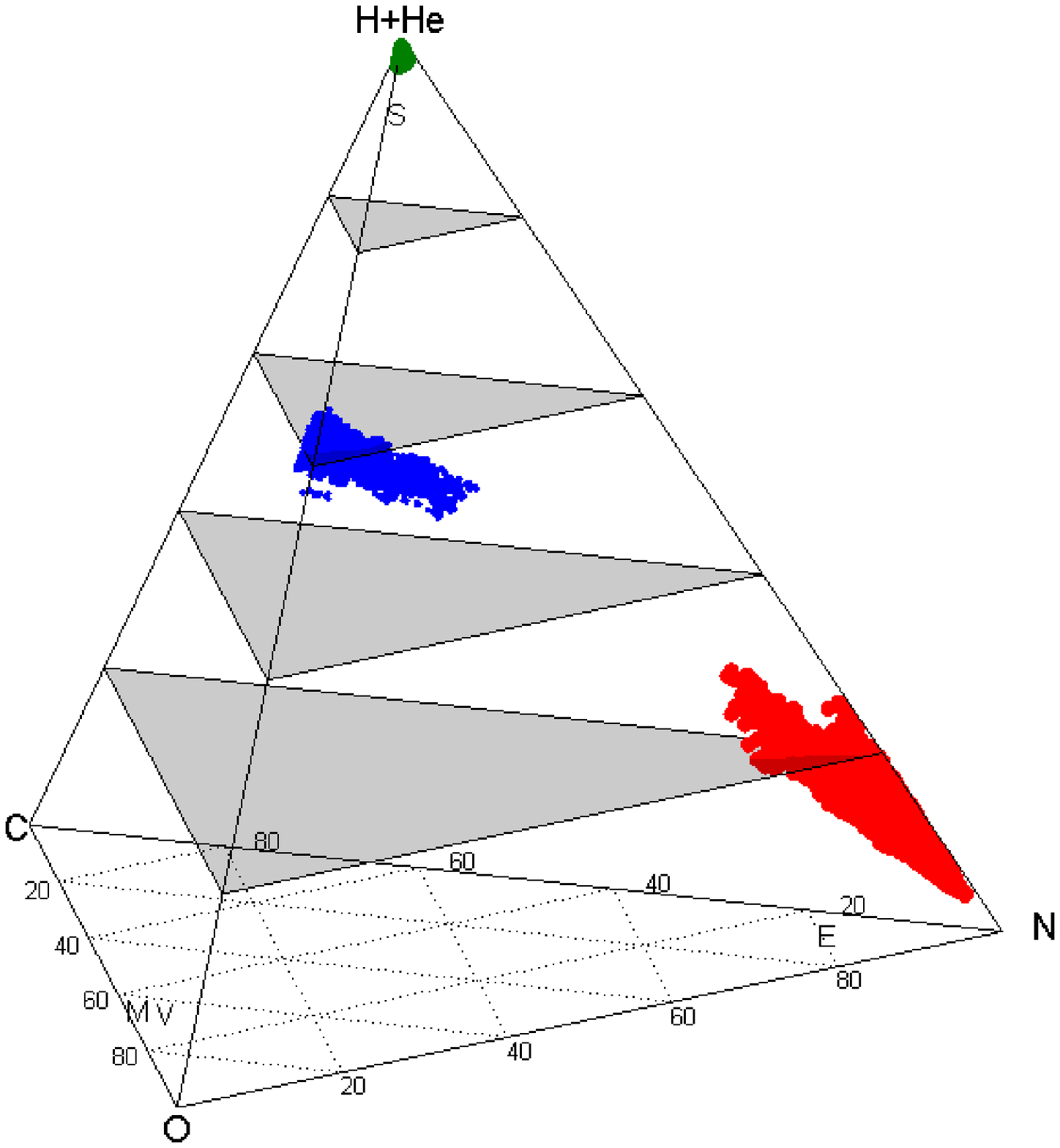}\caption{Quaternary diagram illustrating the posterior probability distributions
for the relative abundances of the elements H, C, O, and N. The colored
volumes represent the 2$\sigma$ Bayesian credible regions of the
elemental composition for the \textquotedbl{}hot Halley world\textquotedbl{}
(blue), the \textquotedbl{}hot nitrogen-rich world\textquotedbl{}
(red) and the \textquotedbl{}hot mini-Neptune\textquotedbl{} (green).
The symbols E, M, V, and S indicate the elemental abundances in the
atmospheres of the Solar System planets Earth, Mars, Venus and Saturn,
respectively, for comparison. The four vertices of the diagram represent
an atmosphere that is fully composed of H, C, O, or N. The opposing
faces are surfaces on which the fraction of H, C, O, or N is zero.
At each point inside the tetrahedron, the elemental fraction is given
by the distances perpendicular to the faces. \label{fig:Constraints-on-elemental}}
\end{figure}

For quantitative constraints, we return to our three scenarios for
hot super-Earth atmospheres. The transmission spectra can clearly
discriminate the different relative abundances of the volatile elements
in the three scenarios (Figure \ref{fig:Constraints-on-elemental})
and may be used to probe their formation history and evolution. The
hot mini-Neptune scenario can be identified to have accreted and retained
a primordial atmosphere dominated by hydrogen, similar to gas and
ice giants in our solar system. At the other end of the parameter
space, the elemental composition of the second scenario indicates
an atmospheric composition dominated by nitrogen. The third scenario
shows an atmosphere that has retained some hydrogen in heavier molecular
species.

\subsection{Total Atmospheric Mass}

We find that transmission spectra present an opportunity to determine
a lower limit for the total mass of the atmosphere of extrasolar planets
based purely on observations. Conceptually, the constraints on the
total atmospheric mass are derived from the constraints on the composition
and the surface pressure of the atmosphere. The ability to constrain
mixing ratios of both the absorbing and the spectrally inactive gases
in the atmosphere enables us to constrain the mean molecular mass
and therefore the mass density in the atmosphere as a function of
pressure. Combined with the independent constraints on the surface
pressure, we can integrate the mass density to estimate the total
column density of the atmosphere. Under the assumption of an approximately
uniform bulk composition and surface pressure around the spherical
planet, we therefore obtain a constraint on the total mass of the
atmosphere.

Two fundamental limits prevent one from accurately constraining the
total atmospheric mass. First, the mass determined from transmission
spectra corresponds to the total atmospheric mass above the uppermost
surface (see Figure \ref{fig:H2_Marginalized-posterior-distribut}(f)
for a quantitative example). Second, following the arguments on the
retrieval of the surface pressure (Section \ref{sub:Surface-Pressure}),
we will not be able to detect the uppermost surface explicitly if
the cloud-free part of the atmosphere is sufficiently thick (Figures
\ref{fig:OutputPlot} and \ref{fig:N2Marginalized-posterior-distribut}).
In conclusion, we can always determine a lower limit on the atmospheric
mass once we have detected spectral features, but determining an upper
limit is only possible if the atmosphere is sufficiently thin for
the surface to be detected and an opaque cloud deck can be excluded
from theoretical principles.

\section{Discussion}

\subsection{Obtaining Observational Constraints on Atmospheric Composition}

The objective in the development of the new retrieval methodology
was to remain independent of model assumptions as much as possible
and let the observational data speak for themselves. By not employing
any assumptions on the elemental composition, chemical equilibrium,
or formation and evolution arguments in the retrieval process, our
results remain independent of preconceived ideas for the planet under
investigation. The atmospheric composition is, instead, completely
described by free parameters and no hidden biases or asymmetries favoring
a particular molecular species in the Bayesian prior are introduced.

The main assumptions in our approach are limited to the principles
of radiative transfer in local thermodynamic equilibrium, hydrostatic
equilibrium, and the correctness of the molecular line lists. For
cases in which no secondary eclipse measurements are available, we
added radiative-convective equilibrium to determine a reasonable temperature
structure. However, since the exact temperature profile has a secondary
effect on the transmission spectrum, we find that this temperature
modeling has little effect on the retrieval results we obtain. In
order to reasonably constrain the atmosphere given the limited data
available in the near future, another guideline in the development
was to keep the number of parameters to a minimum, while still ensuring
that the parameters uniquely define the state of the model atmosphere.
In this study, we assigned a single free parameter for the effective
mixing ratio of each molecular species in the atmosphere, effectively
comparing well-mixed atmospheres to the observation (see Section \ref{sub:Well-mixed-Atmosphere}). 

The main advantage of our retrieval approach for super-Earths over
detailed modeling of atmospheric chemistry and dynamic models is that
it provides an opportunity to discover unexpected types of planets
and atmospheres that do not agree with our current understanding of
formation, evolution and atmospheric processes. For example, no self-consistent
atmospheric chemistry model would predict that the atmosphere of a
terrestrial planet like Earth has an $\mathrm{O_{2}}$ mixing ratio
as high as 21\%. Only the direct interpretation of observations can
tell us about the existence of such unusual atmospheric compositions.
The identification of absorption lines of the $\mathrm{O_{2}}$ absorption
without constraining the high mixing ratio would not be a biosignature
because low abundances of $\mathrm{O_{2}}$ can be a result of photochemical
composition.

In this work, we have shown that we can quantitatively constrain the
atmospheric composition based on observations of the transmission
spectrum, even for super-Earth planets for which the composition is
completely unknown a priori. Transmission spectroscopy is a good tool
for retrieval of the composition because the absorber amount and mean
molecular mass are the main drivers determining the features of the
transmission spectrum, while the influence of the unknown temperature
profile is secondary. We also find, however, that the characterization
of super-Earth planets requires considerably more spectral coverage
and precision than the characterization of the hydrogen-dominated
atmospheres of hot Jupiters. This is not only because of the smaller
signal, but also because of a more complex parameter space that can
result in degeneracies.

\subsection{Non-Unique Constraints for Hazy Atmospheres}

Photochemically-produced hazes may have a significant opacity at short
wavelengths and may mask the signature of molecular Rayleigh scattering
if they are present in the upper atmosphere, . While we may still
be able to probe the near-infrared spectrum and identify molecular
absorbers, we will not be able to probe the transit depth offset of
Rayleigh scattering due to \textit{molecular} scattering. Without
making further assumptions, we will, therefore, lose the ability to
constrain the mixing ratio of the molecular species over orders of
magnitude, even for the major constituents of the atmosphere (see
Section \ref{sub:Atmospheric-Composition}).

By measuring either the slope of the Rayleigh scattering signature
or the shapes of molecular absorption features, we will still obtain
information on the scale height of the atmosphere and, therefore,
obtain an estimate on the mean molecular mass (Section \ref{sub:Mean-molecular-mass}).
We will not, however, be able to constrain the total amount of the
spectrally inactive gases. Since the near-infrared spectrum only constrains
the \textit{relative} abundances of the \textit{absorbing} gases (Section
\ref{sub:Abundance-ratios-of}), we can hypothesize different atmospheric
mixtures of $\mbox{H}_{2}$, $\mbox{N}_{2}$, and absorbing gases
in the correct ratiosthat produce nearly identical transmission spectra.
As a result, we obtain a degeneracy that prevents us from constraining
the molecular abundances uniquely.

One assumption that could be made to compensate for the lack of information
is to not consider the simultaneous presence of nitrogen gas, $\mbox{N}_{2}$,
and hydrogen gas, $\mbox{H}_{2}$. In general, however, the simultaneous
presence of $\mbox{N}_{2}$ and $\mbox{H}_{2}$ cannot be excluded,
even though the preferred chemical form of the two elements N and
H in chemical equilibrium is ammonia $\mbox{NH}_{3}$ at a wide range
of temperatures and pressures because the energy barrier for the reaction
is too large due to strong triple bonds in nitrogen molecules.

\subsection{Stratified Atmospheres \label{sub:Well-mixed-Atmosphere}}

Our parameterization of the atmosphere in the retrieval process assumes
a well-mixed atmosphere. Given the limited amount of data available
in the near-future, the motivation for the assumption of well-mixed
atmospheres is to keep the number of free parameters, which make use
of similar information in the spectrum, small. Observations of the
Solar System planets justify the approach because $\gtrsim$95\% of
the gas in each of the Solar System atmospheres is composed of long-lived,
chemically stable species that were mixed by turbulence and diffusion
for a sufficiently long time \citep[see][and reference therein]{lodders_planetary_1998,pater_planetary_2010}.
If exquisite observations become available in the future, however,
it may be useful to extend the parameterization to retrieve compositional
gradients. For some molecular species, such gradients may be identified
as biomarkers caused by sources at or in the planetary surface.

Physical effects that lead to compositional stratifications of the
gaseous species in the Solar System atmospheres are (1) condensation
of gases that condense at pressure and temperature levels encountered
in the atmosphere, or (2) production or destruction of gas by photochemistry
or geology, or (3) variation of chemical equilibrium with altitude
due to the altitude dependencies of pressure and temperature. Changes
of gas concentration with altitude that are caused by condensation,
however, are usually not relevant for our retrieval because transmission
spectroscopy only probes layers above the condensation clouds. Similarly,
strong changes in the chemical equilibrium usually occur at deep levels
in thick envelopes that are unlikely to be probed in transmission.
In addition, the mixing ratios of gases that do vary with altitude
often only vary over less than one order of magnitude, e.g., CO, $\mbox{H}_{2}\mbox{O}$,
$\mbox{SO}_{2}$ in the atmosphere of Venus \citep{hunten_venus_1983}.
Observational data that are less noisy than the synthetic \textit{JWST}
observations considered in Section \ref{sub:Quantitative-constraints-on}
are necessary to robustly detect such gradients because the retrieved
mixing ratios for minor species in the synthetic \textit{JWST} observations
are uncertain to within one order of magnitude, even for well-mixed
atmospheres (Section \ref{sub:Quantitative-constraints-on}). Photochemistry
or surface sources, however, may lead to concentration gradients that
are substantial at pressure levels probed by transmission spectroscopy
(e.g., ozone in Earth's atmosphere) and may justify extensions to
our parameterization in the future.

\begin{figure*}[tb]
\begin{centering}
\includegraphics[clip,width=1\textwidth]{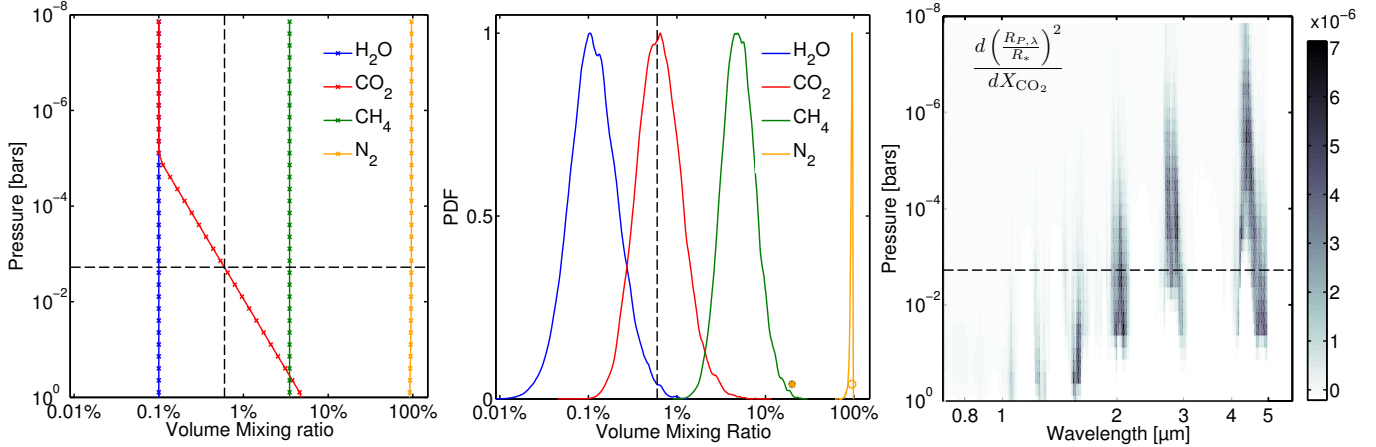}
\par\end{centering}

\caption{Atmospheric retrieval for stratified atmospheres. The left panel illustrates
the volume mixing ratio profiles used to simulate the synthetic \textit{JWST}
observations for a stratified atmosphere scenario. In this scenario,
the volume mixing ratio $X_{\mathrm{CO_{2}}}$ is chosen to decrease
log-linearly from 5\% at 1 bar down to 0.1\% at 0.01 mbar. The error
bars of the observations are similar to ones in Figure \ref{fig:N2Marginalized-posterior-distribut}
(synthetic observations are not shown). The middel panel illustrates
the marginalized posterior probability density as obtained when performing
atmospheric retrieval on the synthetic observations. The abundances
of all molecular species are robustly retrieved. The most likely value
for $X_{\mathrm{CO_{2}}}$ matches the value at the 1-10 mbar level
because the functional derivatives---averaged across the observed
spectrum---are highest for this pressure level (right panel). \label{fig:Atmospheric-retrieval-for}}
\end{figure*}

For atmospheres with a stratified composition, our retrieval method
determines an altitude-averaged mixing ratio that best matches the
observed transmission spectrum. In test cases, we verified that the
atmospheric retrieval method remains robust in providing a reasonable
estimate for the mixing ratios for stratified atmospheres. We simulated
transmission spectra for stratified atmospheres and performed the
retrieval assuming a well-mixed atmosphere (Figure \ref{fig:Atmospheric-retrieval-for}).
We found that, the method remains robust and the retrieved mixing
ratios for stratified gases correspond to the mixing ratios at the
pressure levels at which the functional derivatives with respect to
the mixing ratio are the highest. Using the functional derivatives,
we can, therefore, estimate a posterior at which pressure level we
have probed the atmospheric mixing ratio of the gas. 

\subsection{A Predictive Tool for Planning Observational Programs and Designing
Future Telescopes}

Additional applications of the retrieval method presented here are
(1) to evaluate and optimize observational strategies in the planning
and proposal process of exoplanet observations and (2) to guide the
design of future telescopes and instrumentation for the characterization
of exoplanets. Numerical studies using the retrieval method can provide
concrete guidelines on how many transits must be observed and what
spectral range and spectral resolution is ideal for a specific atmospheric
characterization. The motivation behind the approach is that observational
characterizations of super-Earth atmospheres are extremely challenging
and the observation of many transits with highly capable observatories
will be required.

While the retrieval method is not essential to recognize the need
for higher S/N data than currently available, the retrieval method
is critical to determining exactly what magnitude of data is required
for a useful atmospheric characterization. GJ~1214b is a good example:
In the past, 1 or 2 transits were observed at different wavelengths
by various observers with the goal to characterize the atmosphere
\citep{bean_ground-based_2010,bean_optical_2011,croll_broadband_2011,berta_flat_2012}.
Even though some of these observations approached the theoretical
photon limit, few constraints on the atmosphere could be found. Retrieval
analysis on simulated data shows that ten or more transits are required
with currently available observatories in order to separate out the
two currently most plausible scenarios of a water world and a hydrogen-dominated
atmosphere with high-altitude clouds (B. Benneke et al., in preparation).
We propose a new paradigm in planning observations in which retrieval
analysis of synthetic observations can quantitatively justify the
necessity of large campaigns with ground-based or space-based observatories
for atmospheric characterization.

A conceptually understanding of which details in the spectrum are
required to constrain the atmospheric composition will enable observers
to rationally select the wavelength ranges and spectral resolutions
of transit observations for atmospheric characterization. For example,
constraining the volume mixing ratios of molecular absorbers in super-Earth
atmospheres will require measuring the Rayleigh scattering signature
in addition to the molecule's absorption signatures in the infrared.
If the Rayleigh scattering signature is not observed, even low-noise
observations of the spectral features in the near-infrared with \textit{JWST}
will not provide the information required to determine the volume
mixing ratios.

In addition to the general results, we used simulated \textit{JWST}
NIRSpec observations covering the full range from $\mathrm{0.6}$
to $5\,\text{\ensuremath{\mu}m}$ to show which quantitative constraints
on the composition could be obtained with \textit{JWST} NIRSpec.
An assessment of how well the different atmospheric properties can
be constrained, how many transits are needed, or what observational
parameters are optimal needs to be done with a specific scientific
objective and the available instruments in mind. Therefore, we envision
to use the methodology in the future in collaboration with observers
to evaluate near-future observational opportunities with currently
available instruments.

\subsection{Compositional Retrieval versus Detailed Atmospheric Modeling \label{sub:Compositional-retrieval-as}}

The atmospheric retrieval method and detailed, self-consistent modeling
of a planetary atmosphere \citep{burrows_nongray_1997,seager_dayside_2005,burrows_theoretical_2008}
present two completely complementary approaches to the study of planetary
atmospheres. For studies of solar system planets, it is common practice
to use observational constraints from remote sensing to motivate or
validate detailed modeling of chemistry or dynamics. For a classic
example, the retrieved temperature-pressure profile from radio occultation
measurements on Titan motivated and guided detailed modeling of the
thermal structure to explain the measured temperature profile \citep{mckay_thermal_1989}.
Similarly, the observational detection and abundance constraints on
the methane plume in the Martian atmosphere motivated a multitude
of studies on potential sources and sinks \citep[e.g.,][]{lefevre_observed_2009,krasnopolsky_detection_2004}.

We envision the same kind of complementarity for the characterization
of exoplanets. The strategy would be to use the retrieval method as
presented in this work to identify quantitative constraints on the
atmospheric composition provided by the observations. Then, the constraints
on elemental or molecular abundances can serve as inputs to help guide
the detailed atmospheric modeling to explore physical processes that
would explain the findings. Conversely, atmospheric retrieval can
be complemented by self-consistent forward models in that self-consistent
modeling can further constrain the parameter space by checking the
physical plausibility of the atmospheric scenarios.

\paragraph{\textit{Chemistry-Transport and Photochemistry Models}}

While self-consistent forward models of atmospheric chemistry aim
to provide the physical understanding of the relevant processes in
the atmosphere, self-consistent forward models are dependent on inputs
such as the background atmosphere or elemental abundances, the boundary
conditions at the surfaces, as well as an accurate representation
of all relevant chemical reactions, heat transport and cloud formation
processes. If we knew all inputs and relevant processes a priori,
one could compute the chemical composition and state of the atmosphere
with a self-consistent model. However, many of these inputs will not
be known for exoplanets, especially for planets that do not agree
with our preconceived ideas.

Atmospheric retrieval provides an alternative to self-consistent modeling
forobtaining the composition and state of the atmosphere, but is based
on observations rather than detailed modeling. It can, therefore,
guide the development and application of self-consistent models in
providing constraints on the background atmosphere as well as constraints
on the minor species in atmosphere. As the background atmospheres
of super-Earth planets are not known a priori, it appears that most
self-consistent models of atmospheric chemistry will require that
one uses atmospheric retrieval to infer at least the main species
in the atmospheres of these objects. If the self-consistently modeled
atmospheric properties deviate from the ones provided by the retrieval
analysis of given a set of observations, then the analysis of the
deviation may motivate the inclusion of additional physical or chemical
processes to the self-consistent model (e.g., additional sources and
sinks for molecular species). The combination of self-consistent modeling
and atmospheric retrieval to interpret observations can, therefore,
enhance our understanding of the physical processes in the atmospheres
of extrasolar planets.

\paragraph{\textit{Atmospheric Dynamics}}

One of the most critical factors affecting atmospheric circulation
models is determining the pressure level at which the bulk of the
stellar energy is deposited \citep{heng_influence_2012,perna_effects_2012}.
Atmospheric retrieval may provide a useful input to determine this
pressure level for an observed exoplanet because it allows the constraint
of the molecular composition of the atmosphere, which strongly affects
the opacity of the atmosphere to the incident stellar flux. For hot
Jupiters, previous studies (e.g., \citealp{showman_atmospheric_2009})
assumed chemical equilibrium in combination with solar composition
as a fiducial estimate of the composition. When modeling a specific
planet, the danger is that these assumptions for the composition introduce
inaccuracies in the deposition of stellar light and therefore alter
the results. 

For circulation modeling of super-Earth atmospheres, obtaining observational
constraints on the atmospheric properties is critical. Without observations
or a better understanding of super-Earth planets, even the main constituents
of these atmospheres are unknown, and therefore no fiducial assumptions
on the composition and stellar flux deposition can be made. For rocky
planets, the presence of a solid surface and the pressure level at
the surface play a major role in the atmospheric circulation. The
retrieved surface pressure from observations may, therefore, also
provide an essential input for circulation models.

\section{Summary and Conclusions}

We have presented a Bayesian method to retrieve the atmospheric composition
and thickness of a super-Earth exoplanet from observations of its
transmission spectrum. Our approach is different from previous work
on super-Earths in that we do not test preconceived scenarios, but
retrieve constraints on the atmospheric properties governed by observations,
assuming no prior knowledge of the nature of the planet. Our work
extends previous work on atmospheric retrieval for hot Jupiters in
that we introduce a parameterization that is applicable to general
atmospheres in which hydrogen may not be the dominating gas and clouds
may be present. We infer constraints on individual parameters directly
by marginalizing the joint posterior probability distribution of the
atmospheric parameters. The uncertainty of individual parameters introduced
by complicated, non-Gaussian correlations with other parameters is,
therefore, accounted for in an elegant and straightforward way.

In this work, we have applied the retrieval method to synthetic observations
of the super-Earth GJ~1214b. We investigated which constraints on
the atmospheres of super-Earth exoplanets can be inferred from future
observations of their transmission spectra. Our most significant findings
are summarized as follows.
\begin{itemize}
\item A \textit{unique} constraint of the mixing ratios of the absorbing
gases and up to two spectrally inactive gases is possible with moderate-resolution
($R\sim100)$ transmission spectra, if the spectral coverage and S/N
of the observations are sufficient to quantify (1) the transit depths
in, at least, one absorption feature for each absorbing gas at visible
or near-infrared wavelengths and (2) the slope and strength of the
\textit{molecular} Rayleigh scattering signature at short wavelengths.
Assuming that the atmosphere is wellmixed, and that $\mathrm{N_{2}}$
and a primordial mix of $\mathrm{H_{2}+He}$ are the only significant
spectrally inactive components, one can therefore uniquely constrain
the composition of the atmosphere based on transit observations alone.
\item We can discriminate between a thick, cloud-free atmosphere and an
atmosphere with a surface, where the surface is either the ground
or an opaque cloud deck. For an atmosphere with a surface at low optical
depth, we can quantitatively constrain the pressure at this surface.
A unique constraint of the composition is also possible for an atmosphere
with a surface.\textit{}
\item An estimate of the mean molecular mass made independently of the other
unknown atmospheric parameters is possible by measuring either the
slope of the Rayleigh scattering signature, the shape of individual
absorption features, or the relative transit depths in different features
of the same molecular absorber. For super Earths, discriminating between
hydrogen-rich atmospheres and high mean molecular mass atmospheres
is, therefore, possible, even in the presence of clouds.
\item Determining the volume mixing ratios of the absorbing gases relies
on observations of the molecular Rayleigh scattering signature. Although
the presence of most molecular species can be identified in the near-infrared,
only the relative abundances of the absorbing molecules can be determined
from the infrared spectrum, not their volume mixing ratios in the
atmosphere. The Rayleigh signature of molecular scattering is required
because it enables the measurement of the abundances of spectrally
inactive gases. If the molecular Rayleigh scattering cannot be observed
or is masked by haze scattering at short wavelengths, we will not
be able to determine the volume mixing ratio of the gases in the atmosphere
to within orders of magnitude. The drastic inability to constrain
the mixing ratio was not discovered in previous work on atmospheric
retrieval because hot Jupiters were assumed to be cloud-free and the
mean molecular mass in a hydrogen-dominated atmosphere was known a
priori \citep[e.g.,][]{madhusudhan_temperature_2009}.
\item The retrieval of the mixing ratios of spectrally inactive gases is
fundamentally limited to two independent components. An inherent degeneracy
arises if the atmosphere contains three or more independent spectrally
inactive gases because the same mean molecular mass and the same strength
of the Rayleigh scattering signature can be obtained with different
combinations of the gases.
\item Non-Gaussian treatments of the uncertainties of atmospheric parameters
are essential for atmospheric retrieval from noisy exoplanet observations.
Even given low-noise synthetic observations as considered in this
work, only one-sided bounds and highly non-Gaussian correlations exist
for some atmospheric parameters. Non-Gaussian effects will become
stronger for observational data sets noisier than the synthetic data
considered in this work because the relation between the observables
and the desired atmospheric parameters is highly nonlinear and larger
volumes of the parameter space become compatible with noisier observations.
A limitation of optimum estimation retrieval \citep{lee_optimal_2011,line_information_2012}
for the analysis of noisy exoplanet spectra is, therefore, that the
extent of the confidence regions of atmospheric properties cannot
correctly be described by Gaussian errors around a single best-fitting
solution.
\end{itemize}
Our findings indicate that the retrieval method presented here, combined
with low-noise observations, will provide the opportunity to observationally
characterize atmospheres of individual super-Earth planets and uniquely
identify their molecular and elemental compositions. Similar to observational
constraints on the atmospheres of the Solar System planets obtained
over the last decades, the quantitative constraints obtainable with
our atmospheric retrieval will generally be independent of preconceived
ideas of atmospheric physics and chemistry as well as planet formation
scenarios and atmospheric evolution. The unbiased constraints can,
therefore, motivate the detailed study of the new phenomena in atmospheric
dynamics and chemistry, identify habitability and biosignatures, or
provide clues to planet formation and atmospheric evolution.

We thank Leslie Rogers, David Kipping, Renyu Hu, Julien de Wit, and
Brice Demory for very helpful discussions. We thank Larry Rothman
for access to the HITRAN and HITEMP databases. Support for this work
was provided by NASA.

\appendix{}

\section{An Algebraic Solution to Infer the Mean Molecular Mass \label{sec:An-algebraic-solution}}

For thin or cloudy atmospheres the change in the transit depth across
the spectrum, $\Delta D$, as proposed by \citet{miller-ricci_atmospheric_2009},
cannot be used to uniquely constrain the mean molecular mass because
clouds, hazes, and a surface also affect the feature depths. Here,
we show that measuring the linear slope of the Rayleigh scattering
signature or the shapes of individual features, instead, does provide
constraints on the atmosphere scale height and can be used to estimate
the mean molecular mass for general atmospheres independently of other
atmospheric properties. We derive an algebraic solutions that can
be used to infer the mean molecular mass directly from the transmission
spectrum.

From the geometry described by \citet{brown_transmission_2001}, we
obtain the slant optical depth, $\tau(b)$, as a function of the impact
parameter, $b$, by integrating the opacity through the planet's atmosphere
along the observer's line of sight:

\begin{equation}
\tau_{\lambda}\left(b\right)=2\intop_{b}^{\infty}\sigma_{\lambda}\left(r\right)n\left(r\right)\frac{r\, dr}{\sqrt{r^{2}-b^{2}}}.\label{eq:tau}
\end{equation}

Here, $r$ is the radial distance from the center of the planet. For
Rayleigh scattering, the extinction cross section is only very weakly
dependent on pressure and temperature, and we can assume $\sigma_{\lambda}\left(r\right)=\sigma_{\lambda}$.
Furthermore, motivated by hydrostatic equilibrium, we assume that
the atmospheric number density falls off exponentially according to
$n\left(r\right)=n_{0}e^{-\frac{r}{H}}$, where $H$ is the atmospheric
scale height. With these assumptions we can analytically perform the
integration in Equation (\ref{eq:tau}) and obtain

\begin{equation}
\tau_{\lambda}\left(b\right)=2n_{0}\sigma b\mathcal{K}_{1}\left(\frac{b}{H}\right)\approx2n_{0}\sigma b\sqrt{\frac{\pi}{2\frac{b}{H}}}e^{-\frac{b}{H}},
\end{equation}

where the modified Bessel function of the second kind $\mathcal{K}_{1}\left(x\right)$
is approximated by its asymptotic form $\mathcal{K}_{1}\left(x\right)=\sqrt{\frac{\pi}{2x}}e^{-x}\left[1+O\left(\frac{1}{x}\right)\right]$
for large $x$ \citep{bronstein_taschenbuch_1999}. For spectral regions,
for which the atmosphere is optically thick, the surface does not
affect the transmission spectrum and the observed planet radius as
a function of wavelength can be approximated as

\begin{equation}
R_{P,\lambda}\approx b\left(\tau_{\lambda}=1\right),\label{eq:Rp}
\end{equation}

because the number density falls exponentially with altitude leading
to steep increase in $\tau_{\lambda}$ as a function of $b$. Forming
the ratio between the radii at two different wavelengths, $\lambda_{1}$
and $\lambda_{2}$, for which the extinction cross sections are $\sigma_{1}$
and $\sigma_{2}$, and solving for the scale height, we obtain

\begin{equation}
H|_{r=R_{P}}\approx\frac{R_{P,2}-R_{P,1}}{\ln\left(\frac{\sigma_{2}}{\sigma_{1}}\sqrt{\frac{R_{P,2}}{R_{P,1}}}\right)}\longrightarrow\frac{dR_{P,\lambda}}{d\text{ \ensuremath{\ln\left(\sigma_{\lambda}\sqrt{R_{P,\lambda}}\right)}}}\approx\frac{dR_{P,\lambda}}{d\text{\ensuremath{\left(\ln\sigma_{\lambda}\right)}}},\label{eq:H}
\end{equation}

where we considered the limit of $\lambda_{1}\rightarrow\lambda_{2}$
and then approximated for $\frac{d\ln R_{p,\lambda}}{d\ln\sigma_{\lambda}}$$\ll1$.

Given an estimate of the atmospheric temperature, e.g., $T\approx T_{\mathrm{eq}}$,
we can observationally determine an estimate on the mean molecular
mass

\begin{equation}
\mu_{\mathrm{mix}}=\frac{k_{B}T}{g}\left(\frac{dR_{P,\lambda}}{d\text{\ensuremath{\left(\ln\sigma_{\lambda}\right)}}}\right)^{-1}\times\left(1\pm\frac{\delta T}{T}\right),
\end{equation}

where the factor $\left(1\pm\frac{\delta T}{T}\right)$ accounts for
the inherent uncertainty due to the uncertainty, $\delta T$, in modeling
the atmospheric temperature, $T$, at the planetary radius $r=R_{P,\lambda}$. 

At short wavelengths for which Rayleigh scattering dominates, the
extinction cross section $\sigma$ is proportional to $\lambda^{-4}$,
and we obtain

\begin{equation}
H\approx\frac{R_{P,\lambda_{2}}-R_{P,\lambda_{1}}}{4\ln\left(\frac{\lambda_{1}}{\lambda_{2}}\right)}.\label{eq:H17}
\end{equation}

Given two transit depth observations at $\lambda_{1}$and $\lambda_{2}$
in the Rayleigh scattering regime, we obtain the estimate for the
mean molecular mass

\begin{equation}
\mu_{\mathrm{mix}}=\frac{4k_{B}T}{gR_{*}}\frac{\ln\left(\frac{\lambda_{1}}{\lambda_{2}}\right)}{\left(\frac{R_{p}}{R_{*}}\right)_{\lambda_{2}}-\left(\frac{R_{p}}{R_{*}}\right)_{\lambda_{1}}}\times\left(1\pm\frac{\delta T}{T}\right).
\end{equation}

\end{document}